\documentclass[twocolumn]{aa}
\usepackage{graphicx}
\usepackage{txfonts}
\begin{document}
   \title{The photospheric abundances of active binaries}

   \subtitle{II. Atmospheric parameters and abundance patterns for 6 single-lined RS CVn systems\thanks{Based on observations collected at ESO (La Silla, Chile).}}
   \titlerunning{The photospheric abundances of active binaries II.}

   \author{T. Morel
          \inst{1}
          \and
          G. Micela
          \inst{1}
          \and
          F. Favata
          \inst{2}
          \and
          D. Katz
          \inst{3}
          \and
          I. Pillitteri
          \inst{4}          }

   \offprints{T. Morel,
 \email{morel@astropa.unipa.it}}

   \institute{Istituto Nazionale di Astrofisica, Osservatorio Astronomico di Palermo G.\,S. Vaiana, Piazza del Parlamento 1, I-90134 Palermo, Italy
         \and
             Astrophysics Division - Research and Science Support Department of ESA, ESTEC, Postbus 299, NL-2200 AG Noordwijk, The Netherlands
         \and
             Observatoire de Paris, GEPI, Place Jules Janssen, F-92195 Meudon, France
         \and
             Dipartimento di Scienze Fisiche ed Astronomiche, Universit\`a di Palermo, Piazza del Parlamento 1, I-90134 Palermo, Italy}

   \date{Received ?; accepted ?}

   \abstract{Photospheric parameters and abundances are presented for
     a sample of single-lined chromospherically active binaries from a
     differential LTE analysis of high-resolution spectra. Abundances
     have been derived for 13 chemical species, including several key
     elements such as Li, Mg, and Ca. Two methods have been used. The
     effective temperatures, surface gravities and microturbulent
     velocities were first derived from a fully self-consistent analysis
     of the spectra, whereby the temperature is determined from the
     excitation equilibrium of the \ion{Fe}{i} lines. The second
     approach relies on temperatures derived from the ($B-V$) colour
     index. These two methods give broadly consistent results for the
     stars in our sample, suggesting that the neutral iron lines are
     formed under conditions close to LTE. We discuss the reliability
     in the context of chromospherically active stars of various
     colour indices used as temperature indicators, and conclude that
     the ($V-R$) and ($V-I$) colours are likely to be significantly
     affected by activity processes. Irrespective of the method used,
     our results indicate that the X-ray active binaries studied are
     not as metal poor as previously claimed, but are at most mildly
     iron-depleted relative to the Sun
     ($-0.41\la {\rm[Fe/H]}\la+0.11$). A significant overabundance of
     several chemical species is observed (e.g., the
     $\alpha$-synthezised elements). These abundance patterns are
     discussed in relation to stellar activity.     

   }

   \maketitle
%
%________________________________________________________________

\section{Introduction} \label{sect_intro}
The coronal abundances of active binaries have come under close scrutiny in recent years, thanks to several X-ray satellites such as \emph{XMM-Newton} or \emph{Chandra} (see Favata \& Micela 2003 for a review). While these studies consistently yield a low coronal iron content, the abundance ratios determined for elements with strong X-ray lines (i.e., mainly Fe, Ni, O, Mg, Si, and Ca) appear not only discrepant with respect to the photospheric solar mix, but also significantly different from the values observed in the solar corona, where the abundances of elements with a low first ionization potential are enhanced compared to the solar photosphere (e.g., Audard et al. 2003). 

A detailed comparison between coronal and photospheric abundances in
active binaries would provide useful insights into the chemical
fractionation processes that are likely to operate between the
photosphere and the corona. Unfortunately, such a comparison is made
difficult by the extreme paucity of photospheric abundance
determinations for elements other than iron. Another issue is the
heterogeneous nature of the abundance analyses performed to date,
often resulting in conflicting results in the literature. Most studies
of the iron abundance in active binaries have relied on temperatures
derived from photometric indices (e.g., Randich, Gratton, \&
Pallavicini 1993), despite the fact that 
these stars are known to exhibit photometric anomalies related to
their high level of activity (Gim\'enez et al. 1991; Morale et al.
1996; Favata et al. 1997). The reliability
of photometric colours as temperature indicators in active binaries
and the possible impact on the resulting abundances needs therefore to be addressed. 
In the most
extensive study to date, Randich et al. (1993) and Randich, 
Giampapa, \& Pallavicini (1994) analyzed a total
of 67 components in 54 binary systems by performing a spectral
synthesis over a 25 \AA-wide spectral domain encompassing the
\ion{Li}{i} $\lambda$6708 doublet. The surface gravities were
estimated from the spectral type and the luminosity class, or directly
computed when radii and masses were known. The microturbulent velocity was
calibrated on the luminosity class. The vast majority of these
systems was found to be significantly iron-deficient with respect to
the Sun ($-$1.0 $\la$ [Fe/H] $\la$ +0.2, with a mean of $-$0.4). However,
Fekel \& Balachandran (1993) determined the iron abundance of some
objects in this sample from a spectral synthesis of a similarly small
spectral region, but with temperatures determined by fitting
temperature-sensitive pairs of \ion{Fe}{i} lines, obtaining abundances
up to 0.5 dex higher. Ottmann, Pfeiffer, \& Gehren (1998) determined
temperatures from the wings of the Balmer lines and iron abundances
from a set of \ion{Fe}{ii} lines for a small sample of RS CVn
binaries. At variance with previous investigations, they find
near-solar iron abundances. Abundances of key elements, such as O, Mg,
etc., are only available for very few systems (Donati, Henry, \& Hall
1995; Gehren, Ottmann, \& Reetz 1999; Ottmann et al. 1998; Savanov \& Berdyugina 1994).

To address the above issues we have started a project to derive in a
self-consistent way photospheric parameters and metal abundances for a
large sample of active binaries.  In the first paper of this series
(Katz et al. 2003 --- hereafter Paper I), we have presented a detailed
analysis of \object{HD 113816} (IS Vir) and \object{HD 119285} (V851
Cen), and critically discussed the applicability to active binaries of
three methods commonly used in abundance determinations: (1)
temperature and gravity derived from the excitation and ionization
equilibria of the Fe lines, (2) temperature derived from
colour indices and gravity from the ionization equilibrium of the
Fe lines, and (3) temperature derived from colour indices and
gravity from fitting the wings of collisionally-broadened lines. It
was concluded that these methods gave for both stars broadly
consistent results, although the ($V-I$)-based method gave too low a
temperature. This also argued for small departures from LTE of the
\ion{Fe}{i} lines in the two stars studied. The iron content of
\object{HD 119285} was found to be 0.5 dex higher than the value derived 
by Randich et al. (1994). There was also some indication for a
systematic overabundance of several key elements with respect to the
solar pattern.

In the present paper we extend this study to 6 single-lined systems
and examine their abundance patterns and evolutionary status. We
re-analyze the two stars already discussed in Paper I (\object{HD
  113816} and \object{HD 119285}) because of significant differences
between the 2 line lists; we also derive the abundances of 3
additional chemical species: Li, Cr, and Ba. In addition, we use our
larger sample to re-examine in some more detail the reliability
in the context of active binaries of various methods used in abundance analyses.

\section{Observations and data reduction} \label{sect_obs}
Spectra of 28 active binary systems (18 SB1 and 10 SB2) were acquired
in January 2000 at the ESO 1.52-m telescope (La Silla, Chile) with the
fiber-fed, cross-dispersed echelle spectrograph FEROS in the object+sky configuration. 
The spectral range covered is 3600--9200 \AA, with a resolving power
of 48\,000. 

The program stars are drawn from a
magnitude-limited subset ($V \la 10$) of the catalogue of
chromospherically active stars of Strassmeier et al. (1993). Since the
curve-of-growth abundance analysis presented in the following is unapplicable
to rapid rotators because of severe blending problems, here we restrict
ourselves to studying 6 single-lined systems with a modest
projected rotational velocity ($v \sin i \la 10$ km s$^{-1}$). The
other single-lined systems will be analized by means of other techniques (e.g.,
spectral synthesis), and will be, along with the spectroscopic
binaries, the subject of future publications in this series. Some
basic properties of the stars discussed in this paper are presented in
Table~\ref{tab_obs}. Our sample is made up of stars with very similar
spectral types and is therefore highly homogeneous.

\begin{table*}
\caption{Spectral type, maximum $V$ magnitude, projected rotational velocity, rotational and orbital periods, number of consecutive exposures obtained, mean heliocentric Julian date of the observations, and typical S/N ratio at 6700 \AA \ in the combined spectrum.}
\label{tab_obs}
\begin{center}
\begin{tabular}{llccccccc} \hline
Name                 & Spectral type$^a$ & $V$$^a$ & $v\sin i$$^b$ & $P_{\rm rot}$$^c$ & $P_{\rm orb}$$^c$ & $N$ & ($HJD$--2,451,500) & $S/N$\\
                     &                  & (mag)   & (km s$^{-1}$) & (d)           &   (d)         &                       & \\\hline
\object{HD 10909} (UV For)    & K0 IV            & 8.10    & 2.7           & 64.1          & 30.11         & 2 & 51.59             &150\\
\object{HD 72688} (VX Pyx)    & K0 III           & 6.37    & 7.4           & 19.34         & 45.13         & 1 & 51.80             &250 \\
\object{HD 83442}  (IN Vel)   & K2 IIIp          & 9.10    & 7.5           & 54.95         & 52.27         & 2 & 51.82             &140 \\ 
\object{HD 113816} (IS Vir)   & K2 IV-III        & 8.37    & 5.9           & 24.10         & 23.65         & 3 & 53.79             &180 \\
\object{HD 118238} (V764 Cen) & K2 IIIp          & 9.06    & 11            & 22.62         & 22.74         & 2 & 51.86             &140 \\
\object{HD 119285} (V851 Cen) & K2 IV-III        & 7.69    & 6.5           & 12.05         & 11.99         & 1 & 53.84             &200 \\\hline
\end{tabular}
\begin{flushleft}
$^a$: From Strassmeier et al. (1993).\\
$^b$: Primarily from Fekel (1997) and de Medeiros \& Mayor (1999); from Strassmeier et al. (1993) when not available. For \object{HD 118238}, estimate based on the spectral synthesis of the \ion{Li}{i} $\lambda$6708 doublet (Sect~\ref{sect_lithium}).\\
$^c$: From Strassmeier et al. (1993), except for \object{HD 10909} and \object{HD 113816} (Fekel et al. 2001, 2002).\\
\end{flushleft}
\end{center}
\end{table*}

The data reduction (i.e., bias subtraction, flat-field
correction, removal of scattered light, order extraction and merging,
as well as wavelength calibration) was carried out during the
observations.\footnote{With the {\it ``FEROS Data Reduction
    Software''} available at: {\tt
    http://www.ls.eso.org/lasilla/Telescopes/2p2T/E1p5M/\\FEROS/offline.html}}
Two to three consecutive exposures were generally obtained to allow a
more robust continuum rectification and removal of cosmic ray events.
Changes of physical origin in the spectral characteristics (e.g.,
because of flaring) are very unlikely to operate on the timescales
considered (less than 1200 s), and no significant line-profile
variations between consecutive exposures are indeed found. The journal
of observations is presented in Table~\ref{tab_obs}. The continuum
normalization was achieved in 2 steps. A synthetic Kurucz model
atmosphere (Kurucz 1993) calculated for an initial guess of the
stellar spectral type and luminosity class (with solar metallicity)
was first used to roughly determine the line-free regions, which were
then fitted by low-degree polynomials (the spectra have been
continuum-normalized by segments of 200 to 400 \AA). More accurate
estimates of the atmospheric parameters were obtained after the
spectral analysis detailed below was carried out, and this procedure
was re-iterated.

\section{Line selection and atomic data calibration} \label{sect_lines}

In Paper~I unblended lines were selected from inspection of a
synthetic spectrum computed with Piskunov's {\sc synth} program using
atomic parameters from the VALD database (Piskunov et al. 1995; Kupka
et al. 1999, 2000) and a Kurucz model atmosphere. Here we use
instead a high-resolution spectrum of the K1.5 III star \object{Arcturus}
(Hinkle et al. 2000).  This star is well-matched in terms of spectral
type to the stars under consideration (see Table~\ref{tab_obs}). No
attempts was made to select lines blueward of 5500 \AA \ because of
the difficulty in defining the continuum in this region. Special care
was taken to select a set of neutral iron lines spanning a wide range
in excitation potential and strength.  Low-excitation neutral lines
(except for iron) were discarded, as they are bound to be the most
affected by NLTE effects. We followed the prescription of Ruland
et al.  (1980), by only retaining transitions arising from levels less
than 4.4 eV below the ionization limit. For the few remaining lines
with detailed calculations, the NLTE corrections appear modest: they
range from $\Delta \epsilon=\log(\epsilon)_{\rm
  NLTE}-\log(\epsilon)_{\rm LTE}=-0.09$ to $-0.03$ dex for \ion{Na}{i}
$\lambda$6154, from $+0.05$ to $+0.06$ dex for \ion{Mg}{i}
$\lambda$5711 (Gratton et al.  1999), and from $+0.01$ to $+0.09$ dex
for the Ca lines (Drake 1991). These corrections have been applied to
the abundances derived here.

All $gf$-values have been calibrated on the Sun, i.e., a high S/N
moonlight spectrum acquired with the same instrumental configuration
as the target stars was used in conjunction with a LTE plane-parallel
Kurucz solar model (with $T_{\rm eff}=5777$ K, $\log g=4.44$ cm
s$^{-2}$, and a depth-independent microturbulent velocity $\xi=1.0$ km
s$^{-1}$; Cox 2000) to carry out exactly the same abundance
analysis as detailed below for the active binaries. The oscillator
strengths were then adjusted until Kurucz's solar abundances were
reproduced. For \ion{O}{i} $\lambda$6300.304, we corrected the solar
equivalent width (EW) for the contribution of \ion{Ni}{i} $\lambda$6300.339 (Allende
Prieto, Lambert, \& Asplund 2001). In order to be consistent with
Kurucz models and opacities, we adopt $\log
\epsilon_{\odot}$(Fe)=7.67. As the analysis performed here is purely
differential with respect to the Sun, using the meteoritic value
instead ($\log \epsilon_{\odot}$[Fe]=7.50; Grevesse \& Sauval 1998)
would have no consequence on our results. All lines initially selected
(with the exception of \ion{O}{i} $\lambda$6300) which were too weak
($EW\la10$ m\AA), affected by telluric features, or too distorded to
have their EWs reliably measured in the solar spectrum were discarded.
The derived oscillator strengths were compared with previous estimates in the
literature (Edvardsson et al. 1993; Feltzing \& Gonzalez 2001; Kurucz
\& Bell 1995; Neuforge-Verheecke \& Magain 1997: Reddy et al. 2003),
and 11 lines with a suspicious $gf$-value were rejected. These
exhibited (1) a discrepancy greater than 0.5 dex when the only source
of comparison was Kurucz \& Bell (1995) or (2) a systematic shift
greater than 0.1 dex with other values in the literature.  The final
line list and a discussion regarding the atomic data can be
found in the Appendix.

\section{Methodology} \label{sect_methods}

\hspace*{0.5cm} \emph{Method 1:} The photospheric parameters ($T_{\rm eff}$,
 $\log g$, and $\xi$) and metal
abundances were first determined from a self-consistent analysis of
the spectra by using the measured EWs and the latest generation of
Kurucz LTE plane-parallel atmospheric models computed with the ATLAS9
code (Kurucz 1993) as input for the MOOG software originally developed
by Sneden (1973). We use models with a length of the convective cell
over the pressure scale height, $\alpha=l/H_{\rm p}=0.5$ (e.g.,
Fuhrmann, Axer, \& Gehren 1993), and without overshooting. These
choices have a negligible impact on our derived abundances (less than
0.02 dex). The model parameters ($T_{\rm eff}$, log $g$, $\xi$,
[Fe/H], and [$\alpha$/Fe]) are iteratively modified until: (1) the
\ion{Fe}{i} abundances exhibit no trend with excitation potential or
reduced equivalent width (the EW divided by the wavelength of the
transition), (2) the abundances derived from the \ion{Fe}{i} and
\ion{Fe}{ii} lines are identical, and (3) the Fe and $\alpha$-element
abundances are consistent with the input values. As will be shown in
the following, most of our stars exhibit a noticeable enhancement of
the electron-donor elements Mg, Si, Ca, and Ti. When appropriate, we
therefore used atmospheric models with [$\alpha$/Fe]=0.2 and 0.4 to
account for the increase in the continuous H$^-$ opacity.

A concern is that departures from LTE for the
low-excitation \ion{Fe}{i} lines would bias our determination of the
excitation temperature and ultimately abundances. For K-type
subgiants, the models of Th\'evenin \& Idiard (1999) and Gratton et
al. (1999) suggest NLTE corrections of the order of $\Delta
\epsilon\approx +0.05$ dex. Test calculations for \object{HD 10909}
indicate that artificially decreasing the abundance of the \ion{Fe}{i}
lines with a first excitation potential, $\chi$, below 3.5 eV (Ruland et al.
1980) by this amount would lead to an increase of 85 K and 0.25 dex in
$T_{\rm eff}$ and $\log g$, respectively. The resulting mean iron
abundance would {\em increase} by about 0.06 dex {\em after} fulfilment
  of the excitation and ionization equilibria of the Fe lines. Only
[Ba/Fe] would be significantly affected ($+0.11$ dex), while the
changes would be less than 0.05 dex for the other elements.

\emph{Method 2:} Although the method detailed above appears robust
against NLTE effects at the anticipated level, we further assessed the
reliability of our results by carrying out the same analysis, but after
excluding the \ion{Fe}{i} lines with $\chi$ $<$ 3.5 eV. The effective
temperature was in this case determined from the empirical $T_{\rm
  eff}$-colour calibrations for F0--K5 giants of Alonso, Arribas,
\& Mart\'{\i}nez-Roger (1999, 2001) using the iron abundance obtained
by Method 1.  Table~\ref{tab_photometry} lists the photometric
properties of our program stars, along with the temperatures derived
from the $(B-V)$, $(V-R)$ and $(V-I)$ colours. We use in the following
the temperature determined from the $(B-V)$ index, as being a more
reliable indicator of the stellar effective temperature (Paper I). The
values derived from the $(V-R)$ and $(V-I)$ data will be primarily
used to assess the reliability of colour temperatures in
chromospherically active stars. We do not attempt to use the Balmer
lines to constrain the temperature, as this method is not well-suited
for the coolest stars in our sample (e.g., Barklem et al. 2002).

\begin{table*}
\caption{Distance, extinction in the $V$ band, colour excesses because of 
  interstellar extinction, dereddened colours, and effective
  temperatures. The iron abundances quoted in this table (from Method
  1) were used to derive the effective temperatures from the $(B-V)$
  and $(V-R)$ colours (see Alonso et al. 1999).} 
\label{tab_photometry}
\begin{center}
\begin{tabular}{lcccccc} \hline
 & \object{HD 10909}  & \object{HD 72688}  & \object{HD 83442}  & \object{HD 113816}  & \object{HD 118238} & \object{HD 119285}\\\hline
$d$ (pc)$^a$               & 130   & 131   & 286   & 300   & 760   &   76\\
$A(V)$ (mag)$^b$            & 0.100 & 0.078 & 0.220 & 0.174 & 0.294 & 0.115\\
$E(B-V)$ (mag)$^c$          & 0.030 & 0.024 & 0.067 & 0.053 & 0.090 & 0.035\\
$E(V-R)_{\rm c}$ (mag)$^c$ & 0.019 & 0.015 & 0.041 & 0.033 & 0.055 & 0.022\\ 
$E(V-R)$ (mag)$^c$          & 0.025 & 0.019 & 0.055 & 0.043 & 0.073 & 0.029\\ 
$E(V-I)_{\rm c}$ (mag)$^c$  & 0.040 & 0.031 & 0.089 & 0.070 & 0.118 & 0.046\\ 
$E(V-I)$ (mag)$^c$          & 0.052 & 0.041 & 0.115 & 0.091 & 0.153 & 0.060\\
$(B-V)_0$ (mag)             & 0.930 & 0.926 & 1.103 & 1.057 & 1.170 & 1.045\\
$(V-R)_0$ (mag)             & 0.755 & 0.692 & 0.862 & 0.887 & 0.911 & 0.860\\
$(V-I)_0$ (mag)             & 1.285 & 1.168 & 1.454 & 1.379 & 1.545 & 1.483\\
Reference$^d$               & S93 & C & S93 & HIPP($B-V$), S93 & S93 & C\\
$[$Fe/H$]$                  & --0.41 & +0.11 & +0.02 & --0.11 & --0.12 & --0.23\\\hline
$T_{\rm colour}$($B-V$) (K)    & 4831$\pm$150 & 4989$\pm$150 & 4616$\pm$150 & 4668$\pm$150 & 4470$\pm$150 & 4662$\pm$150\\
$T_{\rm colour}$($V-R$) (K)    & 4741$\pm$100 & 4962$\pm$100 & 4510$\pm$100 & 4445$\pm$100 & 4394$\pm$100 & 4495$\pm$100\\
$T_{\rm colour}$($V-I$) (K)    & 4730$\pm$100 & 4939$\pm$100 & 4469$\pm$100 & 4580$\pm$100 & 4347$\pm$100 & 4429$\pm$100\\
$T_{\rm exc}$ (K)$^e$          & 4830$\pm$87  & 5045$\pm$62  & 4715$\pm$87  & 4700$\pm$77  & 4575$\pm$119 & 4770$\pm$77\\\hline
\end{tabular}
\begin{flushleft}
$^a$: Adopted distance for the determination of the interstellar extinction, derived from \emph{Hipparcos} data (ESA 1997) for the stars with accurate trigonometric parallaxes (see Table~\ref{tab_kinematic}). Distance for \object{HD 118238} from Strassmeier et al. (1993). \\
$^b$: Determined from the empirical model of galactic interstellar extinction of Arenou, Grenon, \& G{\'o}mez (1992). The stellar galactic coordinates are quoted in Table~\ref{tab_kinematic}.\\
$^c$: Conversion factors from Cardelli, Clayton, \& Mathis (1989) and Schlegel, Finkbeiner, \& Davis (1998).\\
$^d$: References for the photometric data. S93: Strassmeier et al. (1993) and references therein; HIPP: \emph{Hipparcos} data (ESA 1997); C: Cutispoto, Messina, \& Rodon{\`o} (2001). The colours in the Cousins system, $(V-R)_{\rm c}$ and $(V-I)_{\rm c}$, were converted into $(V-R)$ and $(V-I)$ Johnson colours following Bessel (1979).\\
$^e$: Temperatures determined from the excitation equilibrium of the \ion{Fe}{i} lines (see text). 
\end{flushleft}
\end{center}
\end{table*}

The uncertainties on the atmospheric parameters arise mainly from the errors inherent
to the determination of the excitation and ionization equilibria of the iron
lines ($T_{\rm eff}$ and $\log g$), or from constraining the abundances given by the \ion{Fe}{i}
lines to be independent of the reduced equivalent width ($\xi$).  To
estimate the uncertainty on $T_{\rm eff}$, for instance, we considered the
1\,$\sigma$ statistical error on the slope of the relation between the
\ion{Fe}{i} abundances and the excitation potentials. The 3 parameters
of the atmospheric models were subsequently varied by the relevant
uncertainty, and the effect on the mean abundance of each element
derived. Other sources of uncertainties include the line-to-line
scatter in the abundance determinations and the errors in the EW
measurements (the latter estimated on a star-to-star basis). These 5
individual ``$1\,\sigma$ errors'' were finally quadratically summed,
ignoring any covariance terms. Following standard practice, this error
budget does not take into account systematic errors. However, the strictly
differential nature of our analysis minimizes many of them (e.g.,
inaccurate atomic data). The abundances of some elements (Na, Mg, and
Ba) were derived from a single, strong line lying on the flat portion
of the curve of growth, and should therefore be treated with some
caution. As in Paper I, we assume for the colour temperatures an
uncertainty of 150 K ($B-V$) and 100 K ($V-R$ and $V-I$). These
figures take into account typical photometric uncertainties, 
internal errors in the calibration, and systematic differences with
other empirical relations (Alonso et al. 1999). For the
metallicity-dependent $T_{\rm eff}$-$(B-V)$ calibration, the
uncertainties on [Fe/H] were also considered.

The EWs (see Table~\ref{tab_ew}) were measured with the task {\it splot} implemented in the
IRAF\footnote{{\tt IRAF} is distributed by the National Optical
  Astronomy Observatories, operated by the Association of Universities
  for Research in Astronomy, Inc., under cooperative agreement with
  the National Science Foundation.} software, assuming Gaussian
profiles. Only lines with a satisfactory fit were
retained. For the most rapid rotators, more severely affected by
blending problems, this resulted in a drastic reduction in the number
of lines used. Measurements performed on consecutive exposures suggest
typical internal errors of the order of 3 m\AA \ ($\approx 4$\%) for
weak or moderately strong lines ($EW \la 100$ m\AA), while this
uncertainty is typically 2 m\AA \ ($\approx 1$\%) for stronger lines.
For \object{HD 113816} and \object{HD 119285}, an excellent agreement
was found for the EWs of the 44 transitions in common with Paper I,
despite a rectification procedure and measurements carried out
completely independently.

Although any abundance analysis should preferentially rely on weak
lines that both lie on the linear part of the curve of growth and are
insensitive to the uncertain treatment of collisional broadening, this
is difficult to achieve in practice for such cool stars with moderate
rotational velocities because of blending problems. As many damping
constants are poorly known, all the lines that may have been
significantly van der Walls broadened (i.e., those with $\log
[EW/\lambda] > - 4.55$) were discarded from the analysis. This damping
process is a minor contributor to the EWs of the remaining
transitions, and was accounted for by the Uns\"old approximation in
the computation of MOOG theoretical curves of growth (Uns\"old 1955).
Calculations including empirical enhancement factors (Simmons \&
Blackwell 1982) would give very similar results ($\Delta T_{\rm eff}
\la 30$ K, $\Delta \log g \la 0.15$ dex, $\Delta {\rm [Fe/H]} \la
0.03$ dex, $\Delta {\rm [M/Fe]} \la 0.05$ dex).  As shown in Paper I,
the use of strong \ion{Fe}{i} lines (with EWs up to 160 m\AA) is not
thought to substantially affect our results. The selected transitions
of the odd-$Z$ elements (Sc, Co, and Ba) are not significantly
broadened by hyperfine structure (Mashonkina et al. 2003; Prochaska et
al. 2000; Reddy et al. 2003). The Sc and Co lines are weak in our
spectra.  Atomic lines affected by telluric features were also
discarded (we used the telluric atlas of Hinkle et al. 2000). Between 37 and 80
lines were used in total for each star (among which between 18 and 47 iron lines).

\section{Results and discussion} \label{sect_results}
\subsection{Abundances and atmospheric parameters} \label{sect_abondance}
Table~\ref{tab_abondance} gives the atmospheric parameters and abundances determined from Methods 1 and 2. Compared to Paper I, the iron abundances of \object{HD 113816} and \object{HD 119285} are lower by 0.15 and 0.10 dex, respectively. The temperatures and surface gravities agree to within 70 K and 0.20 dex, respectively. The abundance ratios of the other chemical elements (prior to our NLTE corrections which were not applied in Paper I) are within the range of the expected uncertainties, with no systematic trends. In some cases (e.g., Mg), the differences are largely due to the different atomic data used.  

The two methods used yield in some cases significantly different values for the effective temperatures and surface gravities (Table~\ref{tab_abondance}). The differences may reach up to 110 K for $T_{\rm eff}$ and up to 0.35 dex for $\log g$. However, this only has a noticeable impact on the abundance of barium (0.13 dex). For the other elements (including Fe), the results given by the 2 methods are not significantly different (0.08 dex at most).

\begin{table*}
\caption{Mean values of the atmospheric parameters and abundances as determined from Methods 1 and 2 ($<>$). The corresponding 1\,$\sigma$ uncertainties and the number of lines included in the abundance analysis ($N$) are also indicated. The abundances of Na, Mg and Ca have been corrected for departures from LTE (see text). The lithium abundances are given in Table~\ref{tab_li}. Blanks indicate that no EWs for the element in question could be reliably measured. We use the usual notation: [A/B]=$\log$ [${\cal N}$(A)/${\cal N}$(B)]$_{\star}$$-$$\log$ [${\cal N}$(A)/${\cal N}$(B)]$_{\sun}$.}
\label{tab_abondance}
\begin{center}
\begin{tabular}{lrrrrrrrrrrrr} \hline
 & \multicolumn{4}{c}{\object{HD 10909} (UV For)} & \multicolumn{4}{c}{\object{HD 72688} (VX Pyx)} & \multicolumn{4}{c}{\object{HD 83442} (IN Vel)}\\
 & \multicolumn{2}{c}{Method 1} & \multicolumn{2}{c}{Method 2} & \multicolumn{2}{c}{Method 1} & \multicolumn{2}{c}{Method 2} & \multicolumn{2}{c}{Method 1} & \multicolumn{2}{c}{Method 2} \\
& \multicolumn{1}{c}{$N$} & \multicolumn{1}{c}{$<>$} & \multicolumn{1}{c}{$N$} & \multicolumn{1}{c}{$<>$} & \multicolumn{1}{c}{$N$} & \multicolumn{1}{c}{$<>$} & \multicolumn{1}{c}{{\em N}} & \multicolumn{1}{c}{$<>$} & \multicolumn{1}{c}{{\em N}} & \multicolumn{1}{c}{$<>$} & \multicolumn{1}{c}{{\em N}} & \multicolumn{1}{c}{$<>$} \\\hline
$T_{\rm eff}$ (K)     &   &   4830$\pm$ 87  &   & 4831$\pm$150    &   &  5045$\pm$62     &   & 4989$\pm$150     &   & 4715$\pm$87      &   &  4616$\pm$150\\
log $g$ (cm s$^{-2}$) &   &  2.90$\pm$0.13  &   & 2.90$\pm$0.13   &   &  2.77$\pm$0.10   &   &  2.58$\pm$0.09   &   &  2.75$\pm$0.13   &   &  2.40$\pm$0.11\\
$\xi$ (km s$^{-1}$)   &   &  1.48$\pm$0.09  &   & 1.48$\pm$0.08   &   &  1.62$\pm$0.06   &   &  1.62$\pm$0.07   &   &  1.89$\pm$0.09   &   &  1.89$\pm$0.10\\
${\rm [Fe/H]}$$^a$        &47 & --0.41$\pm$0.09 &37 & --0.41$\pm$0.12 &42 &    0.11$\pm$0.07 &36 &    0.07$\pm$0.12 &41 &    0.02$\pm$0.10 &35 &  --0.05$\pm$0.14\\
${\rm [Na/Fe]}$   &1  &   0.27$\pm$0.06 &1  &   0.27$\pm$0.10 &1  &    0.45$\pm$0.05 &1  &    0.46$\pm$0.09 &1  &    0.47$\pm$0.06 &1  &    0.48$\pm$0.11\\
${\rm [Mg/Fe]}$   &1  &   0.42$\pm$0.03 &1  &   0.42$\pm$0.08 &1  &  --0.04$\pm$0.04 &1  &  --0.02$\pm$0.09 &1  &    0.13$\pm$0.04 &1  &    0.16$\pm$0.10\\
${\rm [Al/Fe]}$       &2  &   0.42$\pm$0.04 &2  &   0.42$\pm$0.07 &2  &    0.18$\pm$0.06 &2  &    0.20$\pm$0.08 &2  &    0.39$\pm$0.10 &2  &    0.41$\pm$0.14\\
${\rm [Si/Fe]}$       &8  &   0.24$\pm$0.06 &8  &   0.24$\pm$0.08 &4  &    0.10$\pm$0.06 &4  &    0.12$\pm$0.07 &7  &    0.14$\pm$0.15 &7  &    0.19$\pm$0.15\\
${\rm [Ca/Fe]}$   &3  &   0.36$\pm$0.07 &3  &   0.36$\pm$0.15 &3  &    0.08$\pm$0.08 &3  &    0.08$\pm$0.15 &3  &    0.24$\pm$0.06 &3  &    0.22$\pm$0.15\\
${\rm [Sc/Fe]}$       &1  &   0.17$\pm$0.13 &1  &   0.17$\pm$0.19 &   &                  &   &                  &   &                  &   &                 \\
${\rm [Ti/Fe]}$       &1  &   0.35$\pm$0.08 &1  &   0.35$\pm$0.13 &1  &    0.16$\pm$0.09 &1  &    0.16$\pm$0.14 &1  &    0.19$\pm$0.07 &1  &    0.17$\pm$0.13\\
${\rm [Cr/Fe]}$       &4  &   0.07$\pm$0.10 &4  &   0.07$\pm$0.14 &4  &    0.00$\pm$0.08 &4  &   --0.01$\pm$0.14 &2  &    0.15$\pm$0.11 &2  &    0.13$\pm$0.17\\
${\rm [Co/Fe]}$       &1  &   0.18$\pm$0.09 &1  &   0.18$\pm$0.15 &1  &    0.01$\pm$0.10 &1  &    0.00$\pm$0.15 &1  &    0.10$\pm$0.09 &1  &    0.08$\pm$0.15\\
${\rm [Ni/Fe]}$       &10 & --0.02$\pm$0.08 &10 & --0.02$\pm$0.13 &10 &  --0.06$\pm$0.08 &10 &  --0.06$\pm$0.14 &9  &  --0.04$\pm$0.11 &9  &  --0.04$\pm$0.15\\
${\rm [Ba/Fe]}$       &1  & --0.05$\pm$0.11 &1  & --0.05$\pm$0.26 &1  &    0.15$\pm$0.08 &1  &    0.11$\pm$0.24 &1  &    0.10$\pm$0.13 &1  &    0.00$\pm$0.29\\\hline
&&&&&&&&&&&&\\\hline
& \multicolumn{4}{c}{\object{HD 113816} (IS Vir)} & \multicolumn{4}{c}{\object{HD 118238} (V764 Cen)} & \multicolumn{4}{c}{\object{HD 119285} (V851 Cen)}\\
 & \multicolumn{2}{c}{Method 1} & \multicolumn{2}{c}{Method 2} & \multicolumn{2}{c}{Method 1} & \multicolumn{2}{c}{Method 2} & \multicolumn{2}{c}{Method 1} & \multicolumn{2}{c}{Method 2} \\
& \multicolumn{1}{c}{$N$} & \multicolumn{1}{c}{$<>$} & \multicolumn{1}{c}{$N$} & \multicolumn{1}{c}{$<>$} & \multicolumn{1}{c}{$N$} & \multicolumn{1}{c}{$<>$} & \multicolumn{1}{c}{$N$} & \multicolumn{1}{c}{$<>$} & \multicolumn{1}{c}{$N$} & \multicolumn{1}{c}{$<>$} & \multicolumn{1}{c}{$N$} & \multicolumn{1}{c}{$<>$} \\\hline
$T_{\rm eff}$ (K)     &   & 4700$\pm$77      &   &  4668$\pm$150    &   &  4575$\pm$119    &   &  4470$\pm$150    &   &  4770$\pm$77     & &  4662$\pm$150\\
log $g$ (cm s$^{-2}$) &   &  2.45$\pm$0.21   &   &  2.30$\pm$0.15   &   &  2.40$\pm$0.28   &   &  2.08$\pm$0.28   &   &  3.10$\pm$0.19   & &  2.80$\pm$0.21\\
$\xi$ (km s$^{-1}$)   &   &  1.79$\pm$0.08   &   &  1.79$\pm$0.07   &   &  2.78$\pm$0.19   &   &  2.78$\pm$0.21   &   &  1.71$\pm$0.07   & &  1.74$\pm$0.09\\
${\rm [Fe/H]}$$^a$        &46 &  --0.11$\pm$0.09 &40 &  --0.14$\pm$0.13 &20 &  --0.12$\pm$0.13 &18 &  --0.20$\pm$0.16 &44 &  --0.23$\pm$0.10 &36 &  --0.31$\pm$0.13\\
${\rm [Na/Fe]}$   &1  &    0.35$\pm$0.06 &1  &    0.36$\pm$0.11 &1  &    0.41$\pm$0.25 &1  &    0.40$\pm$0.28 &1  &    0.46$\pm$0.05 &1  &    0.47$\pm$0.11\\
${\rm [Mg/Fe]}$   &1  &    0.11$\pm$0.03 &1  &    0.13$\pm$0.09 &   &                  &   &                  &1  &    0.28$\pm$0.03 &1  &    0.31$\pm$0.08\\
${\rm [Al/Fe]}$       &2  &    0.19$\pm$0.06 &2  &    0.21$\pm$0.10 &2  &    0.51$\pm$0.25 &2  &    0.53$\pm$0.29 &2  &    0.48$\pm$0.05 &2  &    0.51$\pm$0.12\\
${\rm [Si/Fe]}$       &4  &    0.08$\pm$0.11 &4  &    0.09$\pm$0.11 &4  &    0.34$\pm$0.18 &4  &    0.39$\pm$0.21 &4  &    0.18$\pm$0.09 &4  &    0.22$\pm$0.11\\
${\rm [Ca/Fe]}$   &3  &    0.25$\pm$0.09 &3  &    0.25$\pm$0.16 &2  &    0.03$\pm$0.15 &2  &  0.01$\pm$0.21 &3  &    0.35$\pm$0.08 &3  &    0.34$\pm$0.17\\
${\rm [Sc/Fe]}$       &1  &  --0.03$\pm$0.13 &1  &  --0.06$\pm$0.19 &   &                  &   &                  &1  &    0.16$\pm$0.16 &1  &    0.09$\pm$0.21\\
${\rm [Ti/Fe]}$       &1  &    0.16$\pm$0.08 &1  &    0.16$\pm$0.14 &1  &    0.36$\pm$0.16 &1  &    0.34$\pm$0.19 &1  &    0.33$\pm$0.08 &1  &    0.32$\pm$0.14\\
${\rm [Cr/Fe]}$       &3  &    0.12$\pm$0.08 &3  &    0.12$\pm$0.15 &1  &    0.09$\pm$0.32 &1  &    0.07$\pm$0.35 &3  &    0.25$\pm$0.15 &3  &    0.24$\pm$0.20\\
${\rm [Co/Fe]}$       &1  &    0.00$\pm$0.08 &1  &  --0.01$\pm$0.14 &1  &    0.11$\pm$0.27 &1  &    0.09$\pm$0.29 &1  &    0.00$\pm$0.10 &1  &  --0.03$\pm$0.14\\
${\rm [Ni/Fe]}$       &9  &  --0.13$\pm$0.10 &9  &  --0.13$\pm$0.14 &6  &  --0.14$\pm$0.22 &6  &  --0.15$\pm$0.25 &9  &  --0.07$\pm$0.10 &9  &  --0.07$\pm$0.15\\
${\rm [Ba/Fe]}$       &1  &    0.34$\pm$0.11 &1  &    0.31$\pm$0.26 &1  &  --0.29$\pm$0.28 &1  &  --0.42$\pm$0.37 &1  &    0.18$\pm$0.14 &1  &  0.07$\pm$0.29\\\hline
\end{tabular}
\begin{flushleft}
$^a$: Mean of the values for \ion{Fe}{i} and \ion{Fe}{ii}.
\end{flushleft}
\end{center}
\end{table*}

Our results regarding the atmospheric parameters and iron content of
these active binaries appear at variance with previous studies and suggest
that these systems may not be as iron-depleted as previously claimed
(Randich et al. 1993, 1994). With the exception of \object{HD 10909},
we derive systematically higher effective temperatures and iron
abundances (see Table~\ref{tab_literature}). The surface gravities are
also significantly different. In the most extreme case (\object{HD
  113816}), we derive a temperature higher by 250 K, a gravity lower
by 1.25 dex and an iron content higher by almost 0.8 dex than 
Randich et al. (1994). These large discrepancies in the iron
abundances are unlikely to be solely accounted for by differences in
the temperature scale; differences in the gravity and microturbulent
velocity (which were often simply calibrated on the spectral type and
luminosity class) are also likely to play a role.

\begin{table*}
\caption{Comparison with previous studies.}
\label{tab_literature}
\begin{center}
\begin{tabular}{lccccccccr} \hline
 & \multicolumn{3}{c}{Fekel \& Balachandran (1993)} &  \multicolumn{3}{c}{Randich et al. (1993, 1994)} &  \multicolumn{3}{c}{This study$^a$}\\
Name     & $T_{\rm eff}$ & $\log g$ & [Fe/H] & $T_{\rm eff}$ & $\log g$ & [Fe/H] & $T_{\rm eff}$ & $\log g$ & \multicolumn{1}{c}{[Fe/H]} \\
     & (K) & (cm s$^{-2}$) & & (K) & (cm s$^{-2}$) & &  (K) & (cm s$^{-2}$) & \\\hline
\object{HD 10909} (UV For)    &      &     &      & 4900 & 3.3 & --0.3 & 4830$\pm$87 & 2.90$\pm$0.13 &  --0.41$\pm$0.09\\
\object{HD 72688} (VX Pyx)    & 4900 & 3.0 & --0.07$^b$ & 4900 & 2.5 & --0.3 & 5045$\pm$62 & 2.77$\pm$0.10 &    0.11$\pm$0.07\\
\object{HD 83442}  (IN Vel)   &      &     &      & 4400 & 2.4 & --0.4 & 4715$\pm$87 & 2.75$\pm$0.13 &    0.02$\pm$0.10\\
\object{HD 113816} (IS Vir)   &      &     &      & 4450 & 3.7 & --0.9 & 4700$\pm$77 & 2.45$\pm$0.21 &  --0.11$\pm$0.09\\
\object{HD 119285} (V851 Cen) &      &     &      & 4650 & 3.6 & --0.6 & 4770$\pm$77 & 3.10$\pm$0.19 &  --0.23$\pm$0.10\\\hline
\end{tabular}
\begin{flushleft}
$^a$ Values derived from Method 1.\\
$^b$ Value rescaled to our adopted solar iron abundance ($\log \epsilon_{\odot}$[Fe]=7.67).
\end{flushleft}
\end{center}
\end{table*}

The oxygen abundance could be
reliably measured from \ion{O}{i} $\lambda$6300 for 2 stars in our sample (\object{HD 10909} and
\object{HD 113816}), but inconsistencies are found with the results given by
the \ion{O}{i} triplet at 7774 \AA. For \object{HD 113816}, for
instance, we obtain ${\rm [O/Fe]}=-0.22$ and $+1.19$ dex for
\ion{O}{i} $\lambda$6300 and the near-IR \ion{O}{i} triplet,
respectively. Such a difference is too large to be accounted for by
granulation effects (Nissen et al. 2002) or by departures from LTE
($\Delta \epsilon = -0.13$ to $-0.05$ dex; Gratton et al. 1999). A
much hotter oxygen line-forming region (perhaps because of
chromospheric heating) would help reducing the discrepancy because of
the inverse sensitivity of these spectral features to the temperature
(e.g., Cavallo, Pilachowski, \& Rebolo 1997). Although this is not clearly
borne out by our exploratory calculations (see below), it remains to
be fully worked out theoretically.  Alternative explanations include
an unusually strong ultraviolet radiation field (Vilhu, Gustafsson, \&
Edvardsson 1987). In view of these problems, the oxygen abundance is
not discussed in the following (the EWs of the oxygen features are
given in Table~\ref{tab_ew} for completeness).

\begin{figure*}
\resizebox{\hsize}{!}
{\rotatebox{0}{\includegraphics{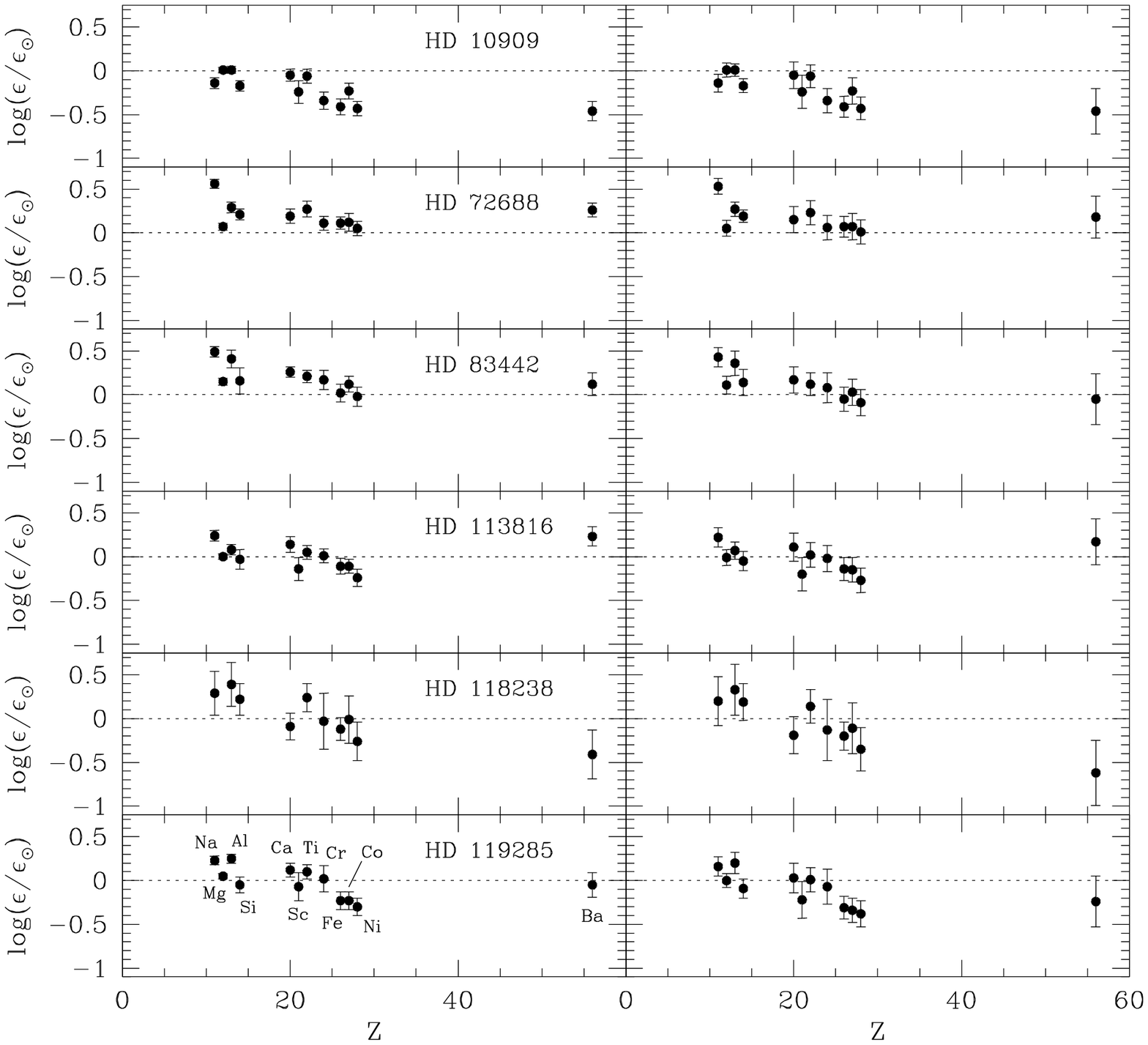}}}
\caption{Abundance patterns for the stars in our sample, determined
  from Method 1 (\emph{left-hand panels}) and Method 2
  (\emph{right-hand panels}).} 
\label{fig_pattern}
\end{figure*}

\begin{figure*}
\resizebox{\hsize}{!}
{\rotatebox{0}{\includegraphics{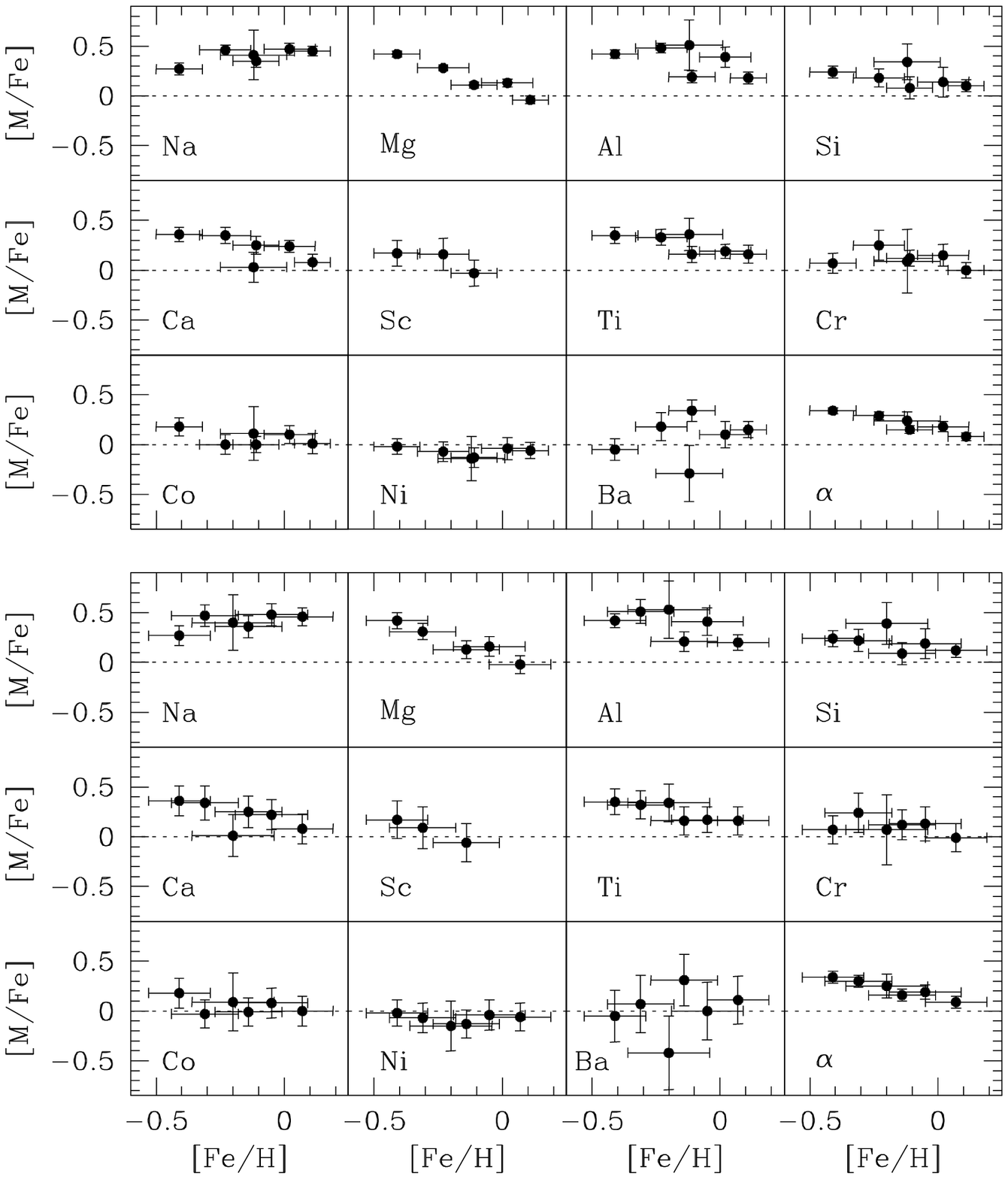}}}
\caption{Abundance ratios as a function of [Fe/H] for Method 1 (\emph{top}) and Method 2 (\emph{bottom}). We define a mean abundance ratio of the  $\alpha$-synthezised elements, [$\alpha$/Fe], as the unweighted mean of the Mg, Si, Ca, and Ti abundances.}
\label{fig_abondance}
\end{figure*}

A striking feature of the abundance
patterns, irrespective of the method used, is the overabundance of most elements with respect to iron
(Fig.~\ref{fig_pattern}).\footnote{It has to be kept in mind that NLTE
  corrections were applied to Na, Mg and Ca, but not to the other
  elements (Sect.~\ref{sect_lines}). There is therefore a systematic, albeit
  small, zero-point offset between [Na/Fe], [Mg/Fe], [Ca/Fe] and the
  other abundance ratios.}  Figure~\ref{fig_abondance} shows the
abundance ratios derived by both methods as a function of [Fe/H].
Only the iron-peak elements Co and Ni present abundances that are
broadly consistent with the solar values, with a hint that nickel is
slightly depleted. The abundance ratio of the neutron-capture element
Ba exhibits a large spread for a given [Fe/H] value, as generally
found for field dwarfs (e.g., Reddy et al. 2003). There is a clear
indication for an increasing overabundance of the $\alpha$-elements
(defined as the mean of the Mg, Si, Ca, and Ti abundances) with
decreasing [Fe/H].

To assess the robustness of our results we carried out the same
analysis on a high-resolution spectrum of \object{Arcturus} (Hinkle et
al. 2000). We used the same line list and calibrated atomic data as
for the active binaries. Our results are given in
Table~\ref{tab_arcturus} where they are compared with previous
estimates in the literature. Our effective temperature and surface
gravity are at the lower end of the previous estimates, but
are very similar to the values of M\"ackle et al. (1975). There is no
evidence that our analysis systematically yields spuriously high
abundance ratios. Compared to the mean values in the literature, the
abundances of the elements heavier than Na differ on average by only
0.04 dex. Once again, we find a discrepancy between the oxygen
abundances given by \ion{O}{i} $\lambda$6300 and the near-IR triplet
(0.6 dex) which is too large to be simply accounted for by NLTE
effects (Gratton et al. 1999; Takeda 2003). This suggests that the
discrepancy observed in the active binaries may not be entirely
attributed to activity, but also to other physical phenomena
  and/or caveats in the atmospheric models. Inconsistencies between
these two oxygen abundance indicators in metal-poor stars is a
long-standing, but still unresolved issue (e.g., Fulbright \& Johnson 2003).

\begin{table*}
\caption{Comparison between our results for \object{Arcturus} and previous estimates in the literature.}
\label{tab_arcturus}
\begin{center}
\begin{tabular}{lccccccc} \hline
 & \multicolumn{7}{c}{References}\\
                   & This study$^a$ & (1) & (2) & (3) & (4) & (5) & (6) \\\hline
$T_{\rm eff}$ (K)     & 4205    & 4260    & 4490   & 4350   & 4330    & 4300  &\\
log $g$ (cm s$^{-2}$) & 1.05    & 0.90    & 2.01   & 1.8    & 1.5     & 1.50  &\\
$\xi$ (km s$^{-1}$)   & 1.59    & 1.8     & 1.8    &        & 1.5     & 1.7   &\\
${\rm [Fe/H]}$        & --0.72  & --0.70  & --0.56 & --0.51 & --0.38  & --0.5 & --0.5\\
${\rm [O/Fe]}$ (6300) &   0.17  &         & 0.40   & 0.26   &         & 0.4   & 0.40\\
${\rm [O/Fe]}$ (7774) &   0.76  &         &        &        &         & 0.4   &\\
${\rm [Na/Fe]}$       &   0.26  &   0.25  &        &        &         & 0.3   & 0.20\\
${\rm [Mg/Fe]}$       &   0.56  &   0.50  &        &        &   0.24  & 0.4   & 0.40\\
${\rm [Al/Fe]}$       &   0.44  &   0.50  &        &        &         & 0.4   &\\
${\rm [Si/Fe]}$       &   0.46  &   0.35  &        &        &   0.19  & 0.4   & 0.30\\  
${\rm [Ca/Fe]}$       &   0.24  &   0.25  &        &        &   0.08  & 0.3   & 0.20\\  
${\rm [Sc/Fe]}$       &   0.09  &   0.05  &        &        &         & 0.2   & 0.20\\  
${\rm [Ti/Fe]}$       &   0.25  &   0.30  &        &        &   0.22  & 0.3   & 0.25\\  
${\rm [Cr/Fe]}$       & --0.02  &   0.05  &        &        &         & 0.0   &\\
${\rm [Co/Fe]}$       &   0.33  &   0.10  &        &        &         &       &\\
${\rm [Ni/Fe]}$       &   0.03  &   0.00  &        &        &   0.11  & 0.0   & 0.10 \\    
${\rm [Ba/Fe]}$       & --0.27  & --0.25  &        &        &         &       &\\\hline
\end{tabular}
\begin{flushleft}
$^a$: For comparison purposes, no NLTE corrections have been applied to our results.\\
Key to references: (1) M\"ackle et al. (1975); (2) Lambert \& Ries (1981); (3) Kj\ae rgaard et al. (1982); (4) Gratton \& Sneden (1987); (5) Peterson, Dalle Ore, \& Kurucz (1993); (6) Donati et al. (1995). 
\end{flushleft}
\end{center}
\end{table*}

\begin{figure}[h!]
\resizebox{\hsize}{!}
{\rotatebox{0}{\includegraphics{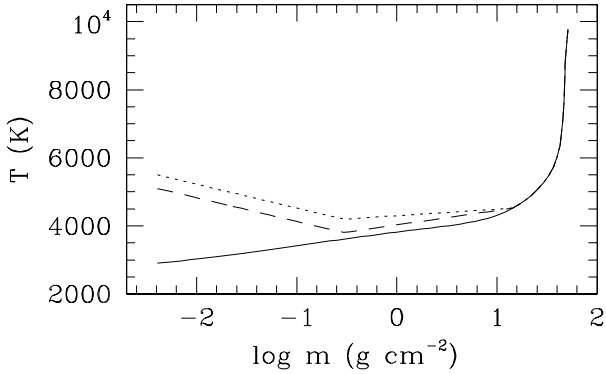}}}
\caption{Temperature structures for \object{HD 10909} of the adopted Kurucz
  model ({\em solid}) and the 2
  empirical models with an added chromospheric component
 ({\em dashed} and {\em dotted} line for $T_{\rm
    min}/T_{\rm eff}=0.788$ and 0.870,
  respectively).}
\label{fig_chromo}
\end{figure}

We now turn to examine the potential importance of chromospheric
heating in inducing the overabundances observed. To this end, we
follow the method of McWilliam et al. (1995), by merging the empirical
chromospheric model of the K0 giant Pollux determined by Kelch et al.
(1978) with our derived photospheric model for \object{HD 10909}. This
modifies the temperature structure of the Kurucz model in two ways:
(1) a reversal of the temperature gradient in the uppermost
photospheric layers corresponding to column mass densities, $m < 0.30$
g cm$^{-2}$. The temperature at this point (which defines the onset of
the chromosphere), $T_{\rm min}$, is set to $T_{\rm min}=0.788 \times
T_{\rm eff}=3805$ K; (2) an overall heating (up to 190 K) of the upper
photosphere between $m = 0.30$ and 14 g cm$^{-2}$. The temperatures of
the Kurucz and empirical models are taken to be identical at $m=14$ g
cm$^{-2}$. The temperature-mass density relation of the merged model
was interpolated between this point and the temperature minimum.  The
electronic density was calculated following Mihalas (1978). We also
experimented with a more prominent chromospheric component, perhaps
more akin to RS CVn binaries (Lanzafame, Bus\`a, \& Rodon\`o 2000). In
this case, $T_{\rm min}=0.870 \times T_{\rm eff}=4200$ K and the
temperature at the outermost layer of the Kurucz model is set to 5500
K (instead of 5100 K). The temperature structure of the 2 atmospheric
models with an added chromospheric component is shown in
Fig.~\ref{fig_chromo}.  Determining an empirical model on a
star-to-star basis (e.g., from fitting the chromospheric emission
component of optical lines) would be obviously preferable, but is
beyond the scope of this paper. We used these 2 atmospheric models to
perform on the EWs determined for \object{HD 10909} the same spectral
analysis as described previously, varying $T_{\rm eff}$, $\log g$ and
$\xi$ until convergence. The chromospheric component was rescaled
during this procedure, with $T_{\rm min}$/$T_{\rm eff}$ held constant
and the temperature at the top of the model chromosphere decreased by
the same amount as the effective temperature. The
differences with the results using the Kurucz models are presented in
Table~\ref{tab_chromo}.  Strong lines are found to be most affected by
the inclusion of a chromosphere, as expected for lines
formed in comparatively higher photospheric layers. The strong
\ion{Fe}{i} lines yield much higher abundances (up to 1.0 dex). As a
result, requiring the iron abundance to be independent of the line
strength leads to a strong increase of the microturbulent velocity.
The overall heating induced by the chromospheric component needs to be
compensated by an ``incipient'' phototospheric model with a much lower
effective temperature (up to 360 K). Achieving ionization equilibrium
of the Fe lines also requires a dramatically lower gravity, with
differences reaching up to 1.1 dex for the model with $T_{\rm
  min}/T_{\rm eff}=0.870$. The fair agreement between our surface
gravities and values derived from evolutionary tracks
(Fig.~\ref{fig_g}) suggests that the effect of the chromosphere on
the photospheric temperature structure may be less severe in reality
than assumed by this model. In summary, assuming that the atmospheres
of chromospherically active stars can be described by a Kurucz
photospheric model leads to an {\it overestimation} of both the
effective temperature and surface gravity. An overall heating of the
upper photospheric layers because of chromospheric activity has a
limited impact on the abundance ratios (except for Ba) and does not
appear to systematically yield higher values (Table~\ref{tab_chromo}).
 The impact of a chromosphere on our abundance analysis is likely
 to be largely due to differences in the temperature structure of the
 upper photosphere where the LTE approximation is nearly valid. However, because this
 assumption of LTE breaks down in the low-density, high-temperature
 chromospheric regions, we warn the reader of the crudeness of these
 pilot calculations.

\begin{table}
\caption{Effect of chromospheric heating on the atmospheric parameters and abundances of \object{HD 10909}. For each quantity $X$, the results are given in the form: $\Delta$$X$=$X$(photosphere)$-$$X$(photosphere+chromosphere). The results for the model without a chromospheric component (Method 1) are given in Table~\ref{tab_abondance}.}
\label{tab_chromo}
\begin{center}
\begin{tabular}{lrr} \hline
$\Delta$$X$ & \multicolumn{2}{c}{$T_{\rm min}$/$T_{\rm eff}$}\\
    & 0.788 & 0.870\\\hline
$\Delta$$T_{\rm eff}$ (K)     & 200 & 360\\
$\Delta$log $g$ (cm s$^{-2}$) & 0.35 & 1.10\\
$\Delta$$\xi$ (km s$^{-1}$)   & --0.37 & --0.49\\
$\Delta$${\rm [Fe/H]}$        & 0.09 & 0.28\\
$\Delta$${\rm [O/Fe]}$ (6300) & 0.06 & 0.23\\
$\Delta$${\rm [O/Fe]}$ (7774) & --0.28 & --0.86\\
$\Delta$${\rm [Na/Fe]}$       & --0.02 & --0.02\\
$\Delta$${\rm [Mg/Fe]}$       & --0.08 & --0.12\\
$\Delta$${\rm [Al/Fe]}$       & --0.06 & --0.08\\
$\Delta$${\rm [Si/Fe]}$       & --0.06 & --0.13\\
$\Delta$${\rm [Ca/Fe]}$       & --0.05 & 0.04\\
$\Delta$${\rm [Sc/Fe]}$       & 0.04 & 0.16\\
$\Delta$${\rm [Ti/Fe]}$       & 0.01 & 0.02\\
$\Delta$${\rm [Cr/Fe]}$       & --0.01 & 0.05\\
$\Delta$${\rm [Co/Fe]}$       & 0.02 & 0.04\\
$\Delta$${\rm [Ni/Fe]}$       & 0.01 & 0.04\\
$\Delta$${\rm [Ba/Fe]}$       & 0.15 & 0.33\\
$\Delta$[$\alpha$/Fe]         & 0.04 & 0.05\\\hline
\end{tabular}
\end{center}
\end{table}

\begin{figure*}
\resizebox{\hsize}{!}
{\rotatebox{0}{\includegraphics{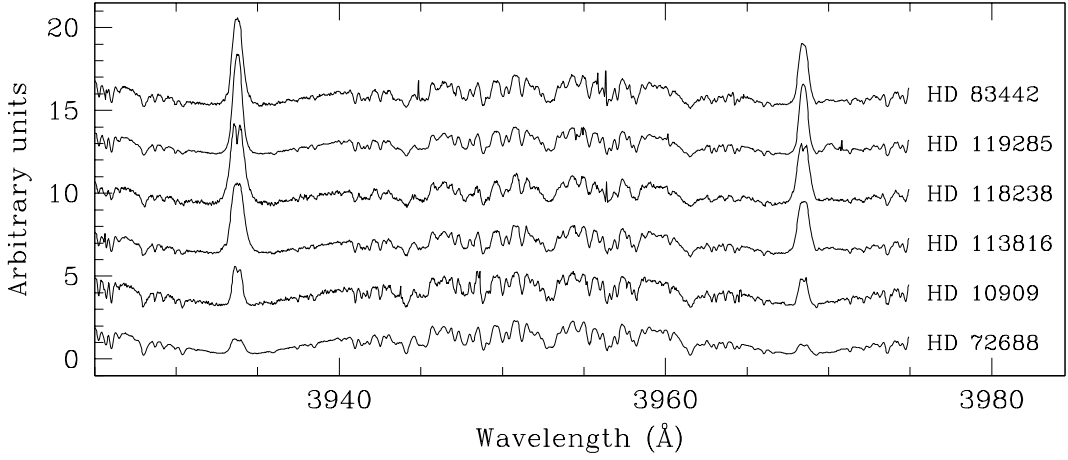}}}
\caption{\ion{Ca}{ii} H+K profiles for our program stars, ordered as a function of decreasing $R_{\rm HK}$ ({\em from top to bottom}; see Table~\ref{tab_index}). The spectra have been normalized to $f_{50}$ (see text) and vertically shifted by a constant value for the sake of clarity. }
\label{fig_ca}
\end{figure*}

To investigate further the effect of stellar activity, we looked for a correlation between
the abundance ratios and the \ion{Ca}{ii} H+K emission-line fluxes (see Fig.~\ref{fig_ca}) calculated following Linsky et al. (1979). The relative fluxes in the \ion{Ca}{ii} lines were first
calculated by integrating the unnormalized spectra in the wavelength
domain $\Delta \lambda({\rm K}_1)=\lambda({\rm K}_{\rm
  1V})-\lambda({\rm K}_{\rm 1R})$ and $\Delta \lambda({\rm
  H}_1)=\lambda({\rm H}_{\rm 1V})-\lambda({\rm H}_{\rm 1R})$ (see
Strassmeier et al. 1990 for definitions), and then dividing these
values by the corresponding flux in the 3925--3975 \AA \ bandpass,
$f_{50}$. All measurements are performed from the zero-flux level.
These relative fluxes, $f$, are converted into absolute surface
fluxes, $\cal{F}$, using (the subscripts refer either to the K or to
the H feature):
\begin{equation}
{\cal F}_{\rm K/H}=50 \: \frac{f_{\rm K/H}}{f_{50}} \: {\cal F}_{50}
\end{equation}
with 
\begin{equation}
\log {\cal F}_{50}=8.264-3.076 \: (V-R)_0 \; \; {\rm for} \; \; (V-R)_0 < 1.3
\end{equation}
This calibration relation is appropriate for the range of colours spanned by our program stars (see Table~\ref{tab_photometry}). The absolute chromospheric line fluxes were subsequently corrected for incipient photospheric contribution:
\begin{equation}
{\cal F}^{'}_{\rm K/H}={\cal F}_{\rm K/H}-{\cal F}^{\rm RE}_{\rm K/H}
\end{equation}
The K and H indices, ${\cal F}^{\rm RE}_{\rm K/H}$, are taken from the radiative equilibrium model atmospheres computed for 3 GK giant stars by Kelch et al. (1978). The ``activity index'', $R_{\rm HK}$, defined as the radiative loss in the \ion{Ca}{ii} H+K lines in units of the bolometric luminosity is given by:
\begin{equation}
R_{\rm HK}=\frac{{\cal F}^{'}_{\rm K}+{\cal F}^{'}_{\rm H}}{\sigma T_{\rm eff}^4}  
\end{equation}
We use the excitation temperature as an estimate of $T_{\rm eff}$ in equation (4), but note that adopting the colour temperatures given by the ($B-V$) index would not significantly affect our results. 

\begin{figure}[h!]
\resizebox{\hsize}{!}
{\includegraphics{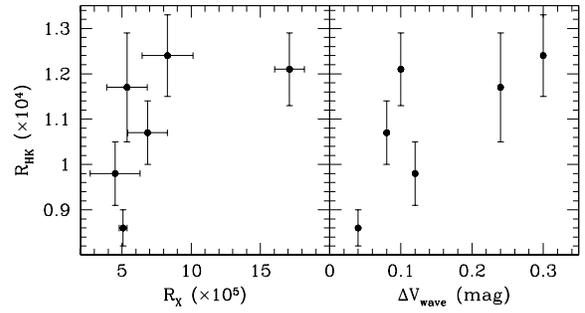}}
\caption{Dependence of $R_{\rm HK}$ on $R_{\rm X}$ ({\em left-hand panel}) and on the maximum amplitude of the wave-like photometric variations in $V$ band ({\em right-hand panel}).}
\label{fig_index}
\end{figure}

\begin{figure*}
\resizebox{\hsize}{!}
{\rotatebox{0}{\includegraphics{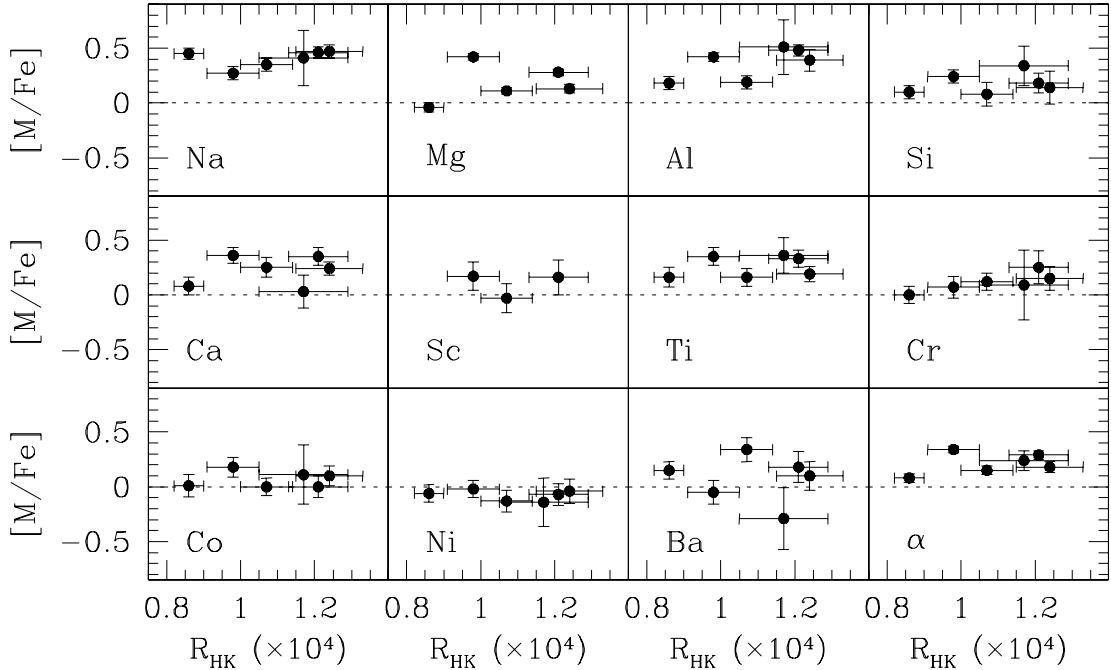}}}
\caption{Abundances (derived from Method 1) as a function of the activity index $R_{\rm HK}$.}
\label{fig_activite}
\end{figure*}

In addition, we define as a ``secondary'' indicator of stellar activity,
$R_{\rm X}$, given as the ratio between the X-ray and the bolometric
luminosities. The latter quantity was derived from evolutionary
tracks (Sect.~\ref{sect_iso}), while the X-ray data are taken from the {\em ROSAT} all-sky
survey (Dempsey et al. 1993, 1997). These ``quiescent'' X-ray
luminosities in the 0.1--2.4 keV energy range have been rescaled to
our more accurate {\em Hipparcos} distances
(Table~\ref{tab_kinematic}).  The major drawback of this activity
indicator is that it may not be representative of the stellar activity
level at the epoch of the spectroscopic observations, a fact which
might contribute to the scatter seen in the plot of $R_{\rm HK}$
against $R_{\rm X}$ (Fig.~\ref{fig_index}). Note that $R_{\rm X}$
  is always well below the empirical threshold defining the
  ``saturation limit'' for the X-ray emission in RS CVn binaries
  ($\sim 10^{-3}$; Dempsey et al. 1993). The
measurements of the activity indices are summarized in
Table~\ref{tab_index}. Finally, we define as an
indicator of stellar spottedness the maximum amplitude of the
wave-like photometric variations in $V$ band, $\Delta V_{\rm wave}$
(values from Strassmeier et al. 1993). Figure \ref{fig_index} suggests
that the incidence of these photospheric features scales (possibly in a non-linear way)
 with the stellar activity level. Figure~\ref{fig_activite} shows the abundance
ratios determined from Method 1 as a function of $R_{\rm HK}$. As
found for the other activity indicators, there is no statistically
significant correlation between the abundance ratios and the stellar
activity level.

\begin{table*}
\caption{Activity indices.}
\label{tab_index}
\begin{center}
\begin{tabular}{lcccccccr} \hline
Name & ($V$--$R$)$_0$ & $T_{\rm eff}$$^a$ & $\log {\cal F}_{\rm K}$ & $\log {\cal F}_{\rm H}$ & $\log {\cal F}^{'}_{\rm K}$ & $\log {\cal F}^{'}_{\rm H}$ & $R_{\rm HK}$$^b$ & \multicolumn{1}{c}{$R_{\rm X}$}\\
     & (mag)         & (K)           & \multicolumn{2}{c}{(erg cm$^{-2}$ s$^{-1}$)} & \multicolumn{2}{c}{(erg cm$^{-2}$ s$^{-1}$)} & ($\times$ 10$^{4}$) & \multicolumn{1}{c}{($\times$ 10$^{5}$)}\\\hline
\object{HD 10909} (UV For)   & 0.755 & 4830$\pm$87 & 6.264 & 6.104 & 6.252 & 6.090 & 0.98$\pm$0.07 & 4.51$\pm$1.80\\
\object{HD 72688} (VX Pyx)   & 0.692& 5045$\pm$62 & 6.297 & 6.116 & 6.281 & 6.095 & 0.86$\pm$0.04 & 5.08$\pm$0.28\\ 
\object{HD 83442}  (IN Vel)  & 0.862& 4715$\pm$87 & 6.331 & 6.139 & 6.326 & 6.133 & 1.24$\pm$0.09 & 8.29$\pm$1.84\\
\object{HD 113816} (IS Vir)  & 0.887& 4700$\pm$77 & 6.254 & 6.077 & 6.249 & 6.071 & 1.07$\pm$0.07 & 6.85$\pm$1.44\\ 
\object{HD 118238} (V764 Cen)& 0.911& 4575$\pm$119 & 6.247 & 6.068 & 6.242 & 6.062 & 1.17$\pm$0.12 & 5.36$\pm$1.46\\ 
\object{HD 119285} (V851 Cen) & 0.860 & 4770$\pm$77 & 6.337 & 6.149 & 6.332 & 6.143 & 1.21$\pm$0.08 & 17.1$\pm$1.07\\\hline
\end{tabular}
\begin{flushleft}
$^a$ Values obtained with Method 1.\\
$^b$ The uncertainties were obtained after propagation of the errors on the absolute emission-line fluxes and effective temperatures.
\end{flushleft}
\end{center}
\end{table*}

\subsection{Discrepancies in the temperature determinations: activity or binarity?} \label{sect_temperature}
The excitation temperatures are found to be systematically
\emph{higher} than the values derived from photometric data
(Table~\ref{tab_photometry}). This disagreement is more severe for
temperatures determined from $(V-R)$ and $(V-I)$ ($\overline{\Delta
  T}\approx 175$ K) than from $(B-V)$ ($\overline{\Delta T}\approx 65$
K).\footnote{Note that interstellar extinction is not a major
    issue here; an uncertainty of 30\% in $A(V)$ would typically
    translate into an error of 30 K in the colour
      temperatures.}

\begin{figure*}
\resizebox{\hsize}{!}
{\includegraphics{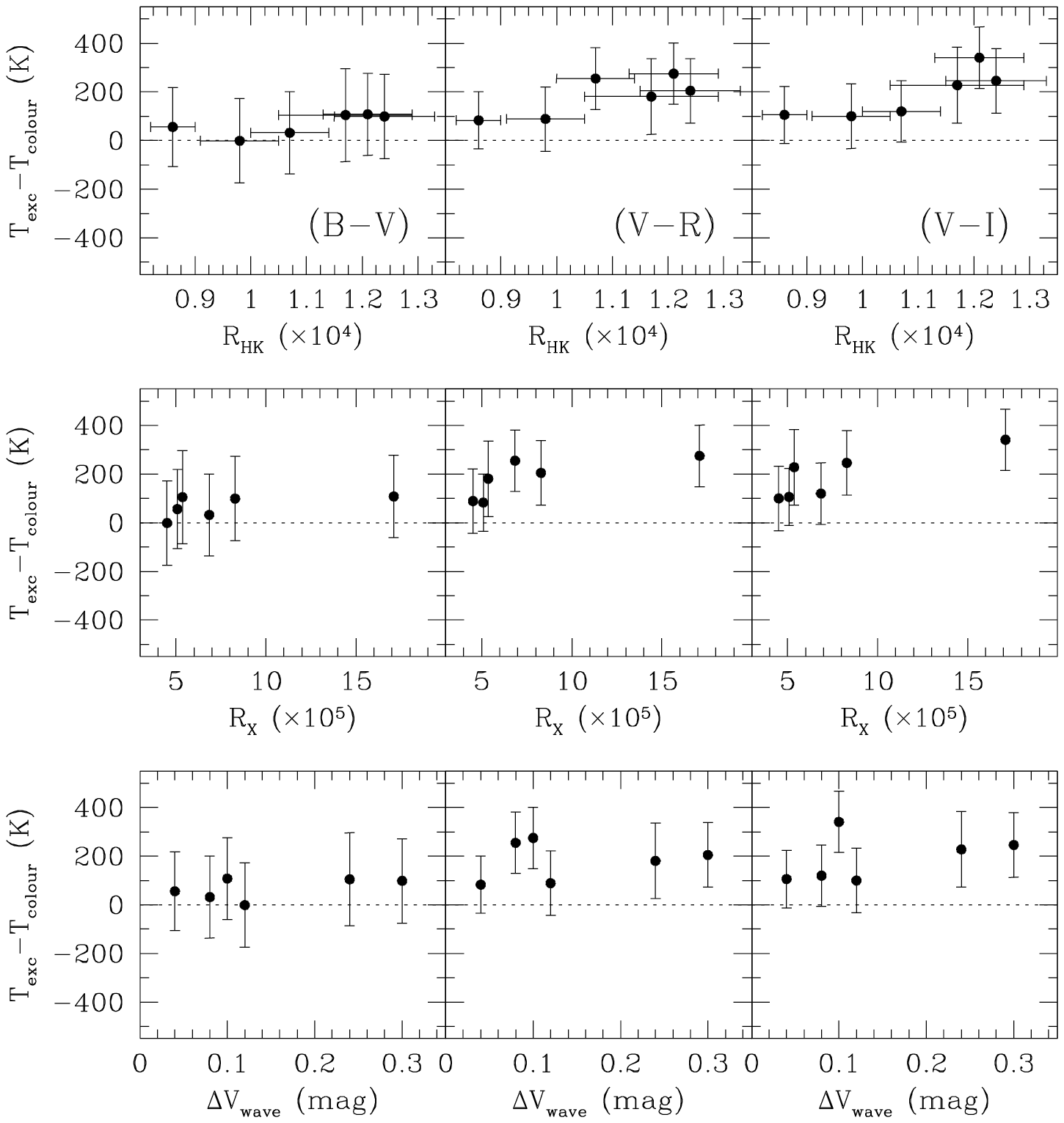}}
\caption{Differences between the excitation and colour temperatures, as a function of the activity indices $R_{\rm HK}$ ({\em top panels}), $R_{\rm X}$ ({\em middle panels}), and maximum amplitude of the wave-like photometric variations in $V$ band, $\Delta V_{\rm wave}$ ({\em bottom panels}). The variations are shown for temperatures derived from ($B-V$) ({\em left-hand panels}), ($V-R$) ({\em middle panels}), and ($V-I$) data ({\em right-hand panels}).}
\label{fig_temperature}
\end{figure*}

Chromospheric activity is likely to affect the emergent photospheric
spectral energy distribution and thus it might induce this discrepancy.
To test this hypothesis, we show in Fig.~\ref{fig_temperature} the
difference between the excitation and colour temperatures, as a
function of $R_{\rm HK}$, $R_{\rm X}$, and $\Delta V_{\rm wave}$. There
is a suggestive indication, in particular from the \ion{Ca}{ii} H+K
data, that the temperature discrepancy increases with the stellar
 activity level. This result is in line with
the recognition that chromospherically active stars exhibit ($V-R$)
and ($V-I$) colour excesses of about 0.06--0.10 mag compared to
otherwise similar, but inactive stars (Fekel, Moffett, \& Henry
1986). On the contrary, the much lower temperature
discrepancy suggested for the ($B-V$) index might be accounted for by
differences in the ($B-V$) vs. $T_{\rm eff}$ calibration chosen (Alonso
et al. 1999). 

The most straightforward explanation for these colour
excesses are photospheric spots, which are ubiquitous in RS CVn
binaries and cover a substantial fraction of the stellar surface (up
to 60\%: O'Neal, Saar, \& Neff 1996). We examined this possibility by
creating a composite, synthetic Kurucz spectrum
which is the sum of two normalized components with $T_{\rm eff}$ =
3830 and 4830 K for the spot and photosphere, respectively. We
  can reasonably assume that the flux emerging from the spots is
  adequately described by a Kurucz model (see, e.g., Solanki \&
  Unruh 1998 in the case of sunspots). We follow the method of
Neff, O'Neal, \& Saar (1995) and consider a single spot lying in the
center of the stellar disk (i.e., no limb-darkening was included) and
with a covering factor, $f_s=30$ or 50\%. We then performed on this
composite spectrum the same analysis as for the active binaries.
We also computed the expected colour excesses (and resulting
temperatures) from the Kurucz models after convolution with the
appropriate filter functions. Table~\ref{tab_spots} gives the colour
and temperature differences compared to the values that would result
for a spotless photosphere. Spots have a roughly similar quantitative
effect on the temperatures derived from spectroscopic and photometric
data, and are unable to explain the distinct behaviour of ($B-V$) and
($V-R$) at high activity levels seen in Fig.~\ref{fig_temperature}.
Although a more detailed treatment is warranted, an interpretation 
of the temperature discrepancies in terms of spots is therefore not
supported by our calculations (see also Fekel et al. 1986). The
 temperature discrepancies are likely to be a
manifestation of activity processes, but the lack of correlation
between the temperature differences and $\Delta V_{\rm wave}$ also suggests
that they may not be strictly related to spots {\em per se}.

\begin{table*}
\caption{Effect of cool spots on the mean colours, as well as on the excitation and colour temperatures.}
\label{tab_spots}
\begin{center}
\begin{tabular}{cccccccc} \hline
$f_s$ & $\Delta$($B-V)_0$ & $\Delta$($V-R)_0$ & $\Delta$($V-I)_0$ & $\Delta T_{\rm exc}$ & $\Delta T$($B-V$) & $\Delta T$($V-R$) & $\Delta T$($V-I$)\\
      & (mag) & (mag) & (mag) & (K) & (K) & (K) & (K) \\\hline
0.3   & 0.026 & 0.022 & 0.065 & 100 &  51 &  60 & 113\\
0.5   & 0.056 & 0.046 & 0.132 & 200 & 109 & 123 & 221\\\hline
\end{tabular}
\end{center}
\end{table*}

A cooler secondary component significantly contributing to the overall
spectral energy distribution would lead to spuriously low colour
  temperatures.\footnote{An enlightening, albeit extreme,
  illustration is provided by \object{HD 101379} (K3 III, GT Mus).
  This system comprises a pair of eclipsing A dwarfs (Murdoch et al.
  1995) and was initially included in our study. It is classified as
  single lined by Strassmeier et al. (1993). From the 3 photometric
  indices, we obtain colour temperatures in the range 5210--5270 K. These
  values are about 650 K higher than the excitation temperature, and
  are clearly inconsistent with the spectral type. Conversely, results
  obtained with Method 1 suggest an iron content (${\rm [Fe/H]}
  \approx -0.45$) difficult to reconcile with the young stellar age
  derived from theoretical isochrones. This low metallicity is likely
  to be spurious, and to result from a dilution of the continuum by
  the hot, early-type eclipsing binary (which is only 0.8 mag fainter
  in $V$).} There are indeed indications for such low-mass companions
for some stars among our sample: \object{HD 113816} (K5 V; Fekel et
al. 2002) and \object{HD 119285} (M3 V; Saar, Nordstr\"om, \& Andersen
1990). The sensitivity of the photometric colours on the nature of the
secondary has been already quantitatively addressed in Paper I.  It
was found that such effects could be substantial for dwarfs, but would
typically lead to differences below 70 K for giants. Regardless of the
seemingly important role played by activity discussed above, binarity
 can therefore hardly account for the large differences
observed in our subgiant stars (up to 340 K).

\subsection{Lithium abundance} \label{sect_lithium}
\begin{figure}[h!]
\resizebox{\hsize}{!}
{\rotatebox{0}{\includegraphics{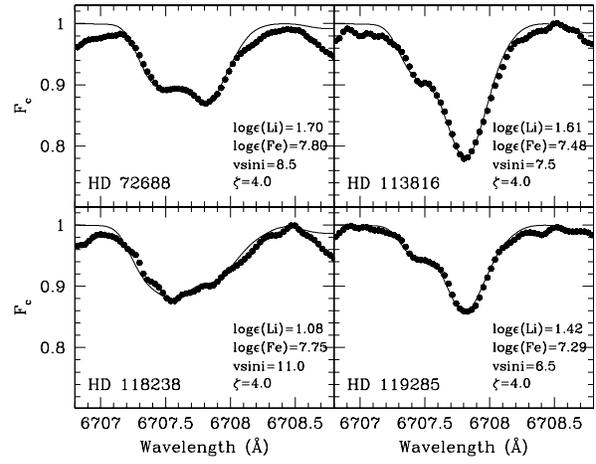}}}
\caption{Spectral synthesis of the \ion{Li}{i} $\lambda$6708 doublet. The atmospheric parameters derived from Method 1 have been used. The resulting lithium and iron abundances (the latter given by \ion{Fe}{i} $\lambda$6707.4), $v\sin i$, and macroturbulent velocity ($\zeta$) are indicated (velocities in km s$^{-1}$).}
\label{fig_li}
\end{figure}

The lithium abundance was determined from a spectral synthesis of the
\ion{Li}{i} $\lambda$6708 doublet. This more sophisticated approach
was made necessary by the blended nature of this doublet at our
instrumental resolution and by its complex atomic structure. The
synthetic spectra were generated using the MOOG software and the line
list of Cunha, Smith, \& Lambert (1995), while the atmospheric
parameters used were those determined previously (Table~\ref{tab_abondance}). 
Because of the
weakness in the solar spectrum of the spectral features of interest, a
relative calibration of the $gf$-values as was done for the other
spectral lines was not attempted. Instead, we use for lithium the
laboratory atomic data quoted in Smith, Lambert, \& Nissen (1998). As
discussed by these authors, the $gf$-values are thought to be
exceptionally accurate (the $^6$Li isotope was not considered). For
the \element[][]{CN} molecules and other atomic lines in the spectral domain of interest, we
adopt the oscillator strengths of Cunha et al. (1995). The projected
rotational velocity, $v \sin i$, along with the iron and lithium
abundances, were adjusted until a satisfactory Gaussian fit of the
blend primarily formed by \ion{Fe}{i} $\lambda$6707.4 and the Li
doublet was achieved (see Fig.~\ref{fig_li}). The fit quality is
largely insensitive to the abundance of the other
chemical species, and solar values were assumed. A very small velocity
shift ($\la$1 km s$^{-1}$) was applied to account for an imperfect
correction of the stellar radial velocity. Mean macroturbulent
velocities, $\zeta$, appropriate to the stellar spectral type and
luminosity class were used (Fekel 1997). Instrumental broadening was
also taken into account. A close agreement was found in all cases
betweeen the iron abundance given by the weak \ion{Fe}{i}
$\lambda$6707.4 line and the mean values in Table~\ref{tab_abondance}. 
Likewise, the $v \sin i$ values
used are in good agreement with previous estimates in the literature
(Table~\ref{tab_obs}). NLTE corrections interpolated from the grids of
Carlsson et al. (1994) for the relevant atmospheric parameters
(typically $\Delta\epsilon=+0.2$ dex) were applied to the lithium
abundances (see Table~\ref{tab_li}). Owing to the weakness of the
\ion{Li}{i} $\lambda$6708 feature in \object{HD 10909} and \object{HD
  83442}, only upper limits could be determined.  Differences between
our abundances (prior to NLTE corrections) and previous estimates in
the literature (Barrado y Navascu\'es et al. 1998; Costa et al. 2002;
Fekel \& Balachandran 1993; Randich et al. 1993, 1994; Soderblom 1985:
Strassmeier et al. 2000) can generally be accounted for by differences
in the temperature scale.

\begin{table}
\caption{NLTE lithium abundances. We use the usual notation: $\log
  \epsilon({\rm Li})=\log {\cal N}({\rm Li})-12.0$. The uncertainties were
  determined as detailed in Sect.~\ref{sect_methods}. } 
\label{tab_li}
\begin{center}
\begin{tabular}{lcc} \hline
Name & \multicolumn{2}{c}{$\log \epsilon$(Li)}\\
     & Method 1 & Method 2\\\hline
\object{HD 10909} (UV For)    & $\la$ 0.6     &  $\la$ 0.6\\
\object{HD 72688} (VX Pyx)    & 1.70$\pm$0.12 & 1.66$\pm$0.19\\ 
\object{HD 83442}  (IN Vel)   & $\la$ 0.6     & $\la$ 0.6\\
\object{HD 113816} (IS Vir)   & 1.61$\pm$0.15 & 1.58$\pm$0.21\\ 
\object{HD 118238} (V764 Cen) & 1.08$\pm$0.20 & 0.92$\pm$0.33\\ 
\object{HD 119285} (V851 Cen) & 1.42$\pm$0.14 & 1.30$\pm$0.21\\\hline
\end{tabular}
\end{center}
\end{table}

\begin{figure}[h!]
\resizebox{\hsize}{!}
{\rotatebox{0}{\includegraphics{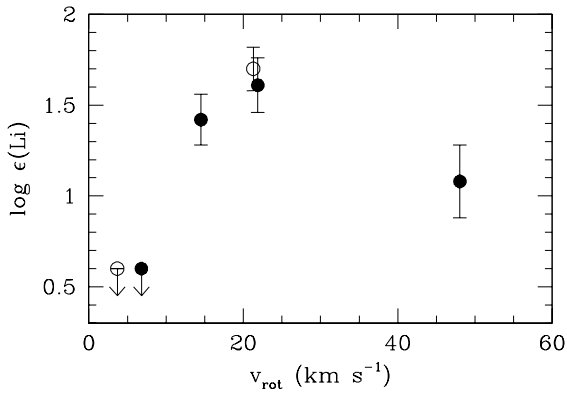}}}
\caption{Lithium abundance as a function of the rotational velocity. The latter quantity was calculated using our effective temperatures (Method 1), the luminosities given by evolutionary tracks (Sect.\ref{sect_iso}), and the rotational periods (Table~\ref{tab_obs}). The synchronous and asynchronous systems are plotted as filled and open circles, respectively.}
\label{fig_li_v}
\end{figure}

Several stars among our sample exhibit a lithium abundance not only higher than what would be expected for such evolved objects, but also higher than single stars with similar characteristics (do Nascimento et al. 2000). Echoing previous investigations, our data do not show any dependence of the Li content on the stellar evolutionary status, mass or activity (e.g., Randich et al. 1993, 1994). The Li abundance is sensitive to a variety of competing physical processes, however, and we feel that meaningful conclusions can only be drawn by seeking statistical trends in much larger samples than analyzed here. We simply note a tentative indication that systems with higher rotational velocities might have retained more of their primordial Li (Fig.~\ref{fig_li_v}; see also Barrado y Navascu\'es et al. 1998). We do not find evidence for systems with synchronized orbits to be less lithium depleted, as would be expected if internal mixing was inhibited in tidally locked binaries (Zahn 1994). This mechanism is perhaps more likely to explain the lack of very lithium-poor stars in close binary systems (Costa et al. 2002).  

\subsection{Stellar Kinematics} \label{sect_kinematic}
\begin{figure}[h!]
\resizebox{\hsize}{!}
{\includegraphics{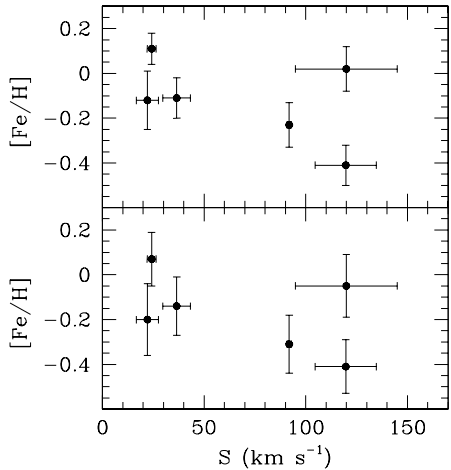}}
\caption{[Fe/H] as a function of the peculiar space velocities for Method 1 (\emph{top}) and Method 2 (\emph{bottom}).}
\label{fig_velocity}
\end{figure}

The space components ($U$, $V$, $W$) were computed from
\emph{Hipparcos} data following Johnson \& Soderblom (1987). They were
subsequently corrected for the effect of differential galactic
rotation (Scheffler \& Els\"asser 1988), by adopting a solar
galactocentric distance of 8.5 kpc and a circular velocity of 220 km
s$^{-1}$.  All kinematic data are quoted in Table~\ref{tab_kinematic} and refer to
the LSR at the star's position and to a right-handed reference system
(i.e., with $\vec{U}$ oriented towards the galactic centre). Rough estimates of the
kinematic ages are also given. A solar
motion $(U, V, W)_{\sun}=(10.0, 5.2, 7.2)$ km s$^{-1}$ was assumed
(Dehnen \& Binney 1998). The peculiar space velocity, $S$, is given
by: $S=(U^2+V^2+W^2)^{1/2}$, and is shown as a function of [Fe/H] in
Fig.~\ref{fig_velocity}. The kinematic properties of \object{HD
  72688}, \object{HD 113816}, and \object{HD 118238} suggest that they
belong to the thin disk population. The situation is less clear for
\object{HD 10909}, \object{HD 83442}, and \object{HD 119285} which
display at least one velocity component typical of the thick disk or
halo (e.g., Soubiran 1993).\footnote{see also: {\tt
    http://physique.obs-besancon.fr/modele/\\descrip.html}.}

\begin{table*}
\caption{Kinematic data, along with kinematic ($\tau_{\rm kin}$;
  Scheffler \& Els\"asser 1988) and evolutionary ages ($\tau_{\rm
    iso}$). No determination of $\tau_{\rm iso}$ from Method 2 was
  possible for \object{HD 119285} (see Fig.~\ref{fig_iso}). The last
  row gives the stellar mass derived from the evolutionary tracks
  (Method 1).} 
\label{tab_kinematic}
\begin{center}
\begin{tabular}{lcccccc} \hline
 & \object{HD 10909} & \object{HD 72688} & \object{HD 83442} & \object{HD 113816} & \object{HD 118238} & \object{HD 119285}\\\hline
$\alpha$ (1950.0) (${\degr}$)      & 26.09 &  127.75 &  143.81 &  195.96 &  203.32 &  205.14\\
$\delta$ (1950.0) (${\degr}$)      & --24.26 &  --34.46 &  --41.80 &  --4.58 &  --33.22 &  --61.12\\
$l$ (${\degr}$)                    & 201.96 &  254.82 &  268.75 &  309.96 &  313.53 &  309.19\\
$b$ (${\degr}$)                    & --77.16 &  3.14 &  7.61 &  57.82 &  28.47 &  0.86 \\
$\pi$ (mas)                        & 7.67$\pm$1.10 &  7.65$\pm$0.59 &  3.50$\pm$1.19 &  3.33$\pm$1.01 &  1.97$\pm$1.19 &  13.13$\pm$1.34\\
$\mu_{\alpha}$ $\cos \delta$ (mas) & 151.03$\pm$1.01 &  --18.26$\pm$0.43 &  --72.49$\pm$0.83 &  --3.19$\pm$1.03 &  --1.98$\pm$1.03 &  22.07$\pm$1.01\\
$\mu_{\delta}$  (mas)              & 97.42$\pm$0.69 &  3.27$\pm$0.47 &  19.53$\pm$0.94 &  --19.23$\pm$0.71 &  --4.43$\pm$0.80 &  16.68$\pm$1.03\\
$v_r$ (km s$^{-1}$)$^a$              & --4.62$\pm$1.65 &  6.40$\pm$4.95 &  48.7$\pm$3.0 &  19.0$\pm$4.0 &  9.3$\pm$3.0 &  93.27$\pm$0.65  \\
$U$ (km s$^{-1}$)                  & --116.7$\pm$15.5 &  --16.6$\pm$1.5 &  --94.4$\pm$30.9 &  8.5$\pm$3.3 &  2.6$\pm$2.9 &  57.4$\pm$0.9\\
$V$ (km s$^{-1}$)                  & --15.9$\pm$1.8 &  --9.6$\pm$4.8 &  --57.5$\pm$3.4 &  --35.1$\pm$7.0 &  --18.9$\pm$5.5 &  --71.8$\pm$1.0\\
$W$ (km s$^{-1}$)                  & 21.1$\pm$3.8 &  --14.7$\pm$0.8 &  --46.5$\pm$15.6 &  --5.3$\pm$5.5 &  --11.1$\pm$5.5 &  --1.4$\pm$0.7\\
$S$ (km s$^{-1}$)                  & 119.7$\pm$15.1 &  24.2$\pm$2.2 &  120.0$\pm$25.1 &  36.5$\pm$6.8 &  22.1$\pm$5.5 &  91.9$\pm$1.0\\
$\tau_{\rm kin}$ (Gyr)             & $\ga$ 9 &  $\sim$0.3 &  $\ga$ 9 &  $\sim$1 &  $\sim$0.2 &  $\sim$9\\
$\tau_{\rm iso}$ (Method 1) (Gyr)  & 7$^{+6}_{-3}$ &  0.6$^{+0.1}_{-0.1}$ &  2.5$^{+3.0}_{-1.5}$ &  1.0$^{+1.5}_{-0.5}$ &  0.3$^{+1.2}_{-0.2}$ &  15$^{+5}_{-5}$\\
$\tau_{\rm iso}$ (Method 2) (Gyr)  & 7$^{+8}_{-4}$ &  0.6$^{+0.2}_{-0.2}$ &  4.0$^{+6.0}_{-2.5}$ &  1.2$^{+1.4}_{-0.6}$ &  0.4$^{+2.6}_{-0.3}$ &  \\
$M$ (M$_{\sun}$)                  & 1.2 &  2.6 &  1.6 &  2.1 &  3.3: &  1.0\\\hline
\end{tabular}
\begin{flushleft}
$^a$ Radial velocities of the center of mass from Strassmeier et al. (1993) and references therein, except for \object{HD 10909} and \object{HD 72688} (de Meideros \& Mayor 1999).
\end{flushleft}
\end{center}
\end{table*}

\subsection{Evolutionary status} \label{sect_iso}
A comparison with theoretical isochrones allows us to examine the
stellar evolutionary status and provides a consistency test on our
derived surface gravities. In virtue of the departures exhibited by
most stars in our sample from a scaled-solar chemical composition
(Fig.~\ref{fig_pattern}), we use the recent set of isochrones for
$\alpha$-element enhanced mixtures of Kim et al. (2002). These
isochrones have been constructed for an $\alpha$-element abundance
($[\alpha/{\rm Fe}]=+0.3$) that is a better match to the overabundances
observed than the value $[\alpha/{\rm Fe}]=+0.5$ used by Salasnich et al.
(2000).  For \object{HD 72688}, we use the isochrones for scaled-solar
mixture of Yi et al. (2001). Both set of evolutionary models use the
same input physics, up-to-date opacities and equations of state, and
are evolved from the pre-main-sequence birthline to the onset of
helium burning in the core.

\begin{figure*}
\resizebox{\hsize}{!}
{\includegraphics{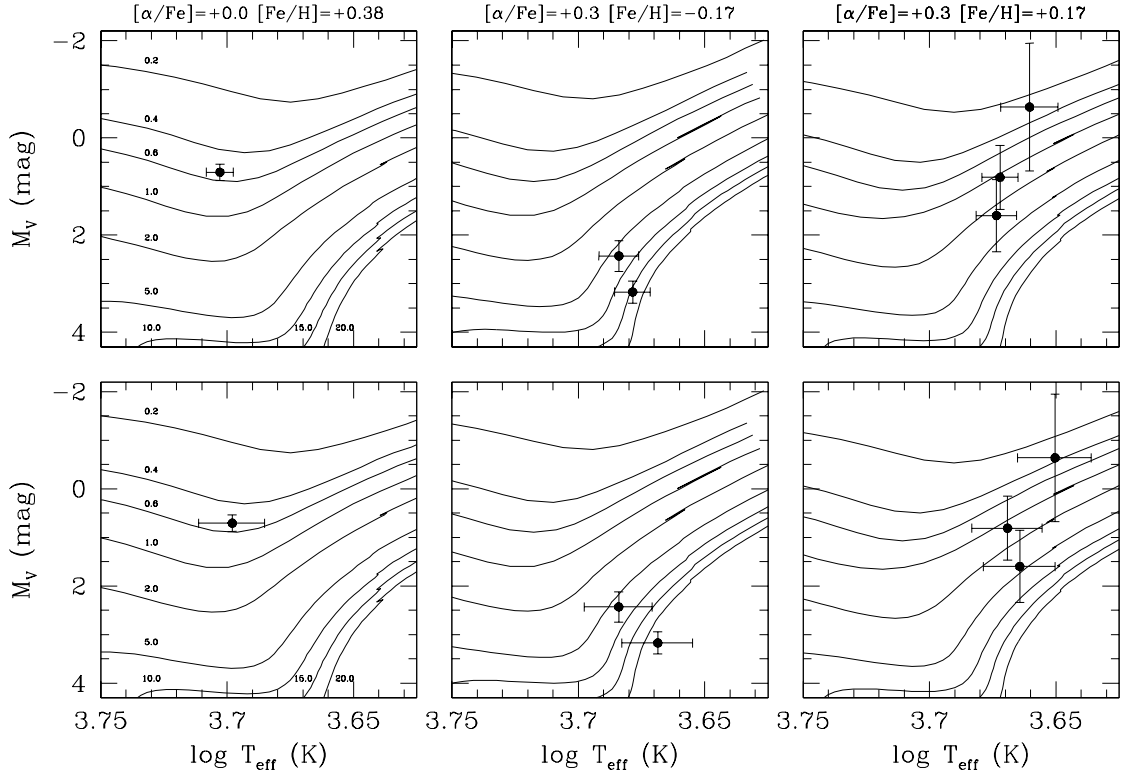}}
\caption{Positions of the program stars in the HR diagrams for Method 1 (\emph{top}) and Method 2 (\emph{bottom}). The theoretical isochrones are shown for [$\alpha$/Fe]=0.0 and [Fe/H]=+0.38 (\emph{left-hand panels}; \object{HD 72688}), [$\alpha$/Fe]=+0.3 and [Fe/H]=$-$0.17 (\emph{middle panels}; from top to bottom: \object{HD 10909} and \object{HD 119285}), as well as [$\alpha$/Fe]=+0.3 and [Fe/H]=+0.17 (\emph{right-hand panels}; from top to bottom: \object{HD 118238}, \object{HD 113816}, and \object{HD 83442}). The metallicity of the isochrones refers to a solar iron abundance: $\log \epsilon_{\odot}$(Fe)=7.50 (Yi et al. 2001; Kim et al. 2002). The age of the isochrones (in Gyr) is indicated in the left-hand panels.}\label{fig_iso}
\end{figure*}

\begin{figure}[h!]
\resizebox{\hsize}{!}
{\includegraphics{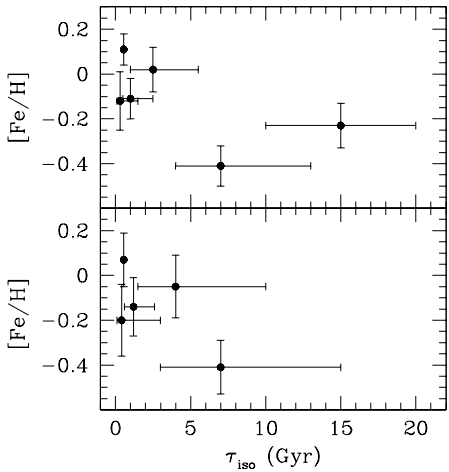}}
\caption{[Fe/H] as a function of the ages derived from the theoretical isochrones for Method 1 (\emph{top}) and Method 2 (\emph{bottom}). For \object{HD 119285}, no age estimate from the evolutionary tracks was possible  from Method 2 (see Fig.~\ref{fig_iso}).}
\label{fig_tauiso}
\end{figure}

The positions of the stars in the HR diagrams for the appropriate Fe
and [$\alpha$/Fe] abundances are shown in Fig.~\ref{fig_iso}. The
metallicity of the isochrone was chosen to match as closely as
possible our spectroscopic values (allowance was made for the
different solar iron abundances assumed). The absolute magnitudes are
not significantly affected by the presence of a faint, main-sequence
companion. Our stars share a very similar evolutionary status and are
all starting to ascend the red giant branch. Estimates of the stellar
masses were derived from the evolutionary tracks (Table
\ref{tab_kinematic}), and suggest that these stars have evolved
  from main-sequence progenitors with masses ranging from about 1.0 to
  3.3 M$_{\sun}$ (i.e., of spectral type G2--A0). The evolutionary
age derived from Method 2 for \object{HD 119285} appears uncomfortably
large and points to an underestimated effective temperature. However,
the loci in the $M_{V}$-$T_{\rm eff}$ plane are very sensitive to the
somewhat subjective choice of the isochrone metallicity.
Furthermore, if HD 119285 belongs to the thick disk (a possibility
that cannot be ruled out from its kinematics;
Table~\ref{tab_kinematic}), then we would expect it to be about 12 Gyr
old. This value is, within the uncertainties, consistent with the
isochrones. The evolutionary ages are in fair agreement with the
kinematic ages (Table~\ref{tab_kinematic}), and are shown as a
function of [Fe/H] in Fig.~\ref{fig_tauiso}.

\begin{figure}[h!]
\resizebox{\hsize}{!}
{\includegraphics{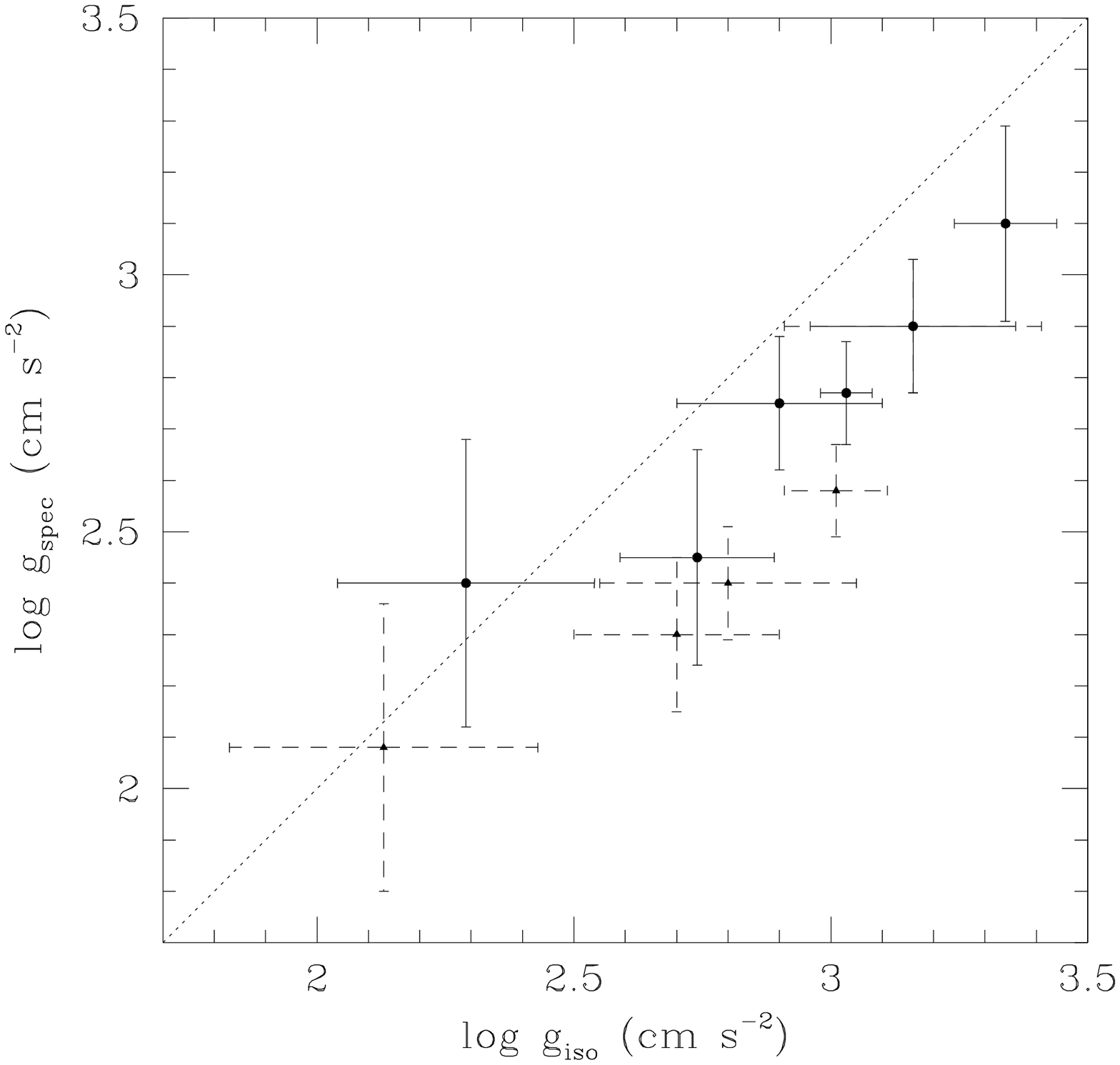}}
\caption{Comparison between the gravities given by the theoretical isochrones and the values derived from Method 1 (\emph{solid}) and  Method 2 (\emph{dashed}). In the latter case, no estimate of the gravity from the evolutionary tracks was possible for \object{HD 119285} (see Fig.~\ref{fig_iso}). The error bars for the theoretical gravities were derived from the uncertainties on the position of the stars in the HR diagrams.}
\label{fig_g}
\end{figure}

A comparison between the gravities derived from the ionization
equilibrium of the Fe lines and from the evolutionary tracks is shown
in Fig.~\ref{fig_g}. A reasonable agreement is found, although there
is a clear indication that the spectroscopic gravities are
systematically lower by about 0.15 dex. Whatever the cause of this
putative discrepancy (see Allende Pietro et al. 1999 for a thorough
discussion), we simply stress here that such a gravity offset would
have little impact on the abundance patterns ($\Delta[{\rm
  M/Fe}]\la0.05$ dex) and would thus not affect the conclusions
presented in this paper.

\section{Conclusions} \label{sect_conclusions}
The temperatures derived from ($V-R$) and ($V-I$) colour indices appear to be affected in our sample by activity processes whose exact nature remains to be identified. On the contrary, a fair agreement between the excitation temperatures and the values derived from the ($B-V$) data is found, suggesting that this colour index might be a more robust indicator of the effective temperature in chromospherically active binaries (provided that the metallicity is known with reasonable accuracy). 

Irrespective of the method used, our study suggests that active binaries may not be as iron-deficient as previously thought (see also Ottmann et al. 1998). The physical mechanisms leading to the noticeable metal enrichment observed remain to be identified, although an interpretation in terms of nucleosynthesis yields remains viable. In particular, the gradient between [$\alpha$/Fe] and [Fe/H] is much steeper than for thin disk dwarfs (e.g., Reddy et al. 2003), but is reminiscent of the thick disk population (e.g., Feltzing, Bensby, \& Lundstr\"om 2003). Part of the gradient could therefore be due in our sample to a mix of the 2 populations (thin and thick disk stars; Sect.~\ref{sect_kinematic}). On the other hand, these stars are presumably not evolved enough to have experienced deep convective mixing (as supported by the high Li abundance in some of them; Table~\ref{tab_li}). The abundance peculiarities observed are thus hard to explain in the framework of standard evolutionary theory. 

Our data do not show a clear correlation between the abundance ratios
and the activity level (Fig.~\ref{fig_activite}), therefore firm
conclusions regarding the potential role played by chromospheric
activity in inducing the overabundances observed must await the
analysis of a much larger sample. Our exploratory calculations
suggest that chromospheric heating of the upper photosphere is
unlikely to significantly affect the abundance patterns. A more
  sophisticated and rigorous approach (e.g., relaxing the assumption
  of LTE) is, however, needed to investigate this issue further and to
  establish the importance of other physical processes causally
  related to activity.

\begin{acknowledgements}
  This research was supported through a European Community Marie Curie
  Fellowship. G.\,M. and I.\,P. acknowledge financial support from ASI
  (Italian Space Agency) and MIUR (Ministero della Istruzione,
  dell'Universit\`a e della Ricerca). We made use of the Simbad
  database operated at CDS, Strasbourg, France. We wish to thank an
  anonymous referee for useful comments.
\end{acknowledgements}

\appendix
\section{Line list and oscillator strengths} \label{sect_ew}
\begin{figure}[h!]
\resizebox{\hsize}{!}
{\rotatebox{0}{\includegraphics{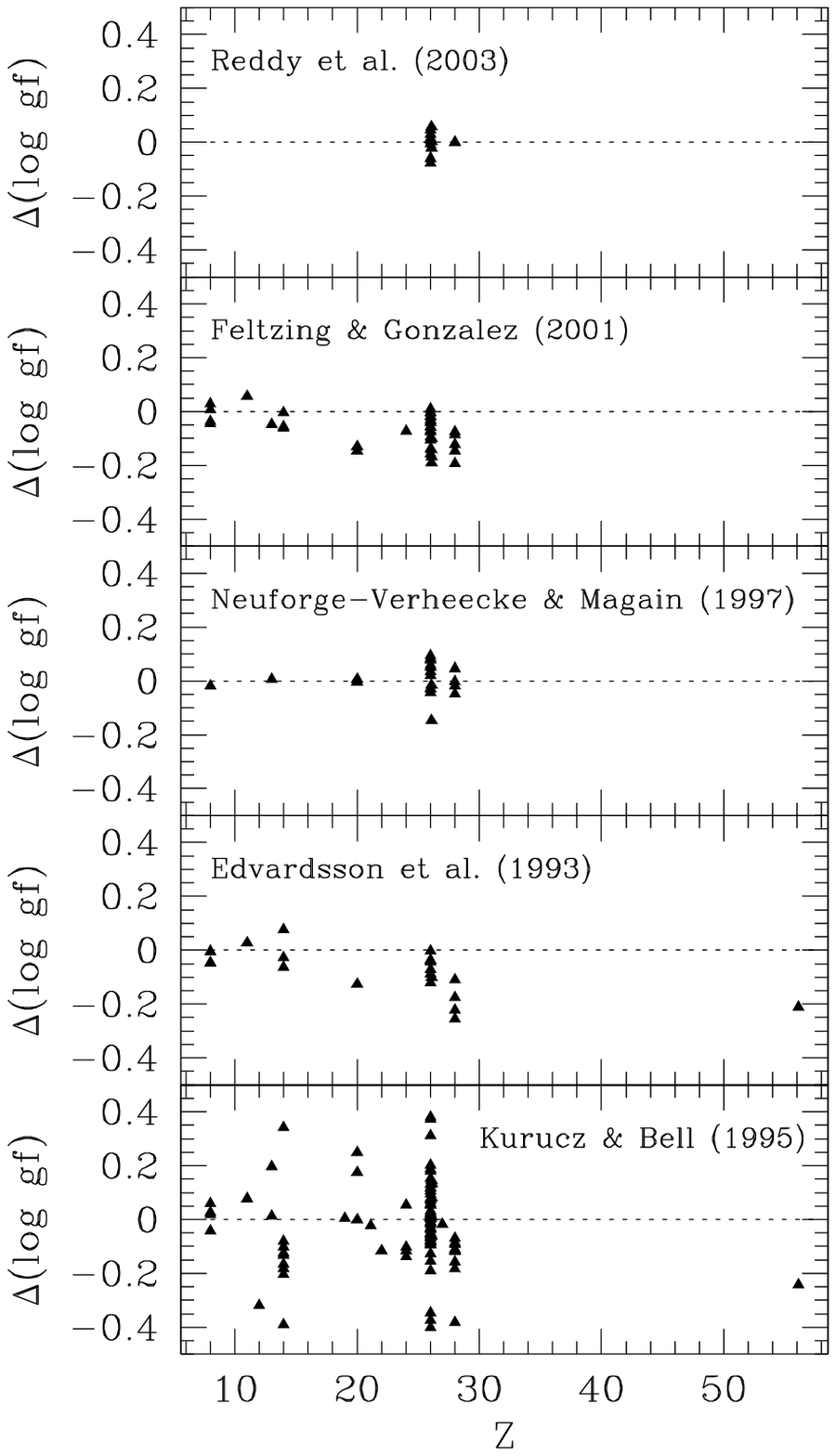}}}
\caption{Comparison between our calibrated $gf$-values and previous
  determinations in the literature. The differences are given in the
  form $\Delta(\log gf)=\log gf({\rm literature})-\log gf({\rm this~
    study})$. The values of Kurucz \& Bell (1995) for \ion{Fe}{i}
  $\lambda$7780 and \ion{Si}{i} $\lambda$6029 are off-scale, and are
  likely erroneous.}
\label{fig_gf}
\end{figure}

Table~\ref{tab_ew} gives for each spectral line the wavelength,
excitation potential (Kurucz \& Bell 1995), the adopted $\log gf$-value
 (see Sect.~\ref{sect_lines}), and the EW measurements. Kurucz
solar abundances are listed along with the element symbols.
Fig.~\ref{fig_gf} presents a comparison between our $gf$-values and
previous determinations in the literature after rescaling to Kurucz
solar abundances (Edvardsson et al. 1993; Feltzing \& Gonzalez 2001;
Kurucz \& Bell 1995; Neuforge-Verheecke \& Magain 1997; Reddy et al.
2003). Except for Kurucz \& Bell (1995), all $gf$-values have also
been calibrated on the Sun. Neuforge-Verheecke \& Magain (1997) values
were determined from the Li\`ege Solar Atlas and by using the solar
model of Holweger \& M\"uller (1974). Feltzing \& Gonzalez (2001)
derived their values from the Kitt Peak Solar Flux Atlas (Kurucz,
Furenlid, \& Brault 1984) and a MARCS solar model (Gustafsson et al.
1975). Edvardsson et al. (1993) and Reddy et al. (2003) used a
purpose-built solar model and an ATLAS9 Kurucz model, respectively.
They both used a solar spectrum obtained as part of their
observational programs. In contrast, the $gf$-values of Kurucz \&
Bell (1995) come from a variety of experimental studies, and are
therefore highly heterogeneous. This is likely to be reflected in the
large scatter seen in Fig.~\ref{fig_gf} ({\em bottom}).

The agreement between our values and those found in the literature
using a similar approach is reasonably good (the differences never
exceed 0.25 dex). The agreement is in particular excellent with
Neuforge-Verheecke \& Magain (1997) and Reddy et al. (2003). However,
a systematic offset for elements heavier than Ca is found with
Edvardsson et al. (1993) and Feltzing \& Gonzalez (2001). There is
some indication for an increasing offset between Edvardsson et al.
(1993) values and ours for stronger lines. This
discrepancy might stem from our neglect of some damping processes in the
determination of the solar abundances (e.g., van der Walls broadening
implemented without enhancement factors), as also done by
Neuforge-Verheecke \& Magain (1997) and Reddy et al. (2003). However, because
our analysis of RS CVn binaries is strictly differential with respect
to the Sun, where lines have comparable strength, this offset
has no bearing on our derived abundances.

\begin{table*}
\caption{Calibrated atomic data and EW measurements}
\label{tab_ew}
\begin{center}
\begin{tabular}{lcrrrrrrr} \hline
  $\lambda$ (\AA) & $\chi$ (eV)  & log $gf$ & \multicolumn{6}{c}{EW (m\AA)$^a$}\\
     &    &          & UV For & VX Pyx & IN Vel & IS Vir & V764 Cen & V851 Cen\\\hline            
{\bf \ion{O}{i}}; $\log$ $\epsilon_{\odot}$(O)=8.93&&&&&&&&\\
  6300.304  & 0.000 & --9.778 &  20.1 &       &       &  23.4 &       &     \\   
  7771.944  & 9.147 &   0.297 &  65.3 &  81.6 &  84.0 &  92.3 & 125.1 &  93.8\\
  7774.166  & 9.147 &   0.114 &  54.4 &       &  75.8 &       & 120.7 &  84.2\\
  7775.388  & 9.147 & --0.064 &  38.7 &  58.3 &       &       &       &  53.9\\
{\bf \ion{Na}{i}}; $\log$ $\epsilon_{\odot}$(Na)=6.33&&&&&&&&\\
  6154.226  & 2.102 & --1.637 &  62.1 & 103.4 & 124.3 & 100.5 & 139.8 &  97.0\\
{\bf \ion{Mg}{i}}; $\log$ $\epsilon_{\odot}$(Mg)=7.49&&&&&&&&\\
  5711.088  & 4.346 & --1.514 & 138.7 & 139.6 & 163.9 & 151.2 &       & 152.5\\
{\bf \ion{Al}{i}}; $\log$ $\epsilon_{\odot}$(Al)=6.47&&&&&&&&\\
  6698.673  & 3.143 & --1.843 &  51.6 &  59.9 &  84.5 &  62.9 &  95.5 &  70.5\\
  7835.309  & 4.022 & --0.663 &  65.0 &  83.7 & 111.0 &  79.8 & 144.0 &  92.4\\
{\bf \ion{Si}{i}}; $\log$ $\epsilon_{\odot}$(Si)=7.55&&&&&&&&\\
  5793.073  & 4.930 & --1.894 &  49.2 &  75.6 &  73.2 &  61.3 &  86.4 &  56.0\\
  5948.541  & 5.083 & --1.098 &  88.0 &       &       &       &       &      \\
  6029.869  & 5.984 & --1.553 &       &       &  38.2 &       &  42.2 &      \\
  6155.134  & 5.620 & --0.742 &  71.5 & 105.0 &  91.3 &  83.9 & 106.4 &  74.6\\
  6721.848  & 5.863 & --1.100 &  41.3 &  71.6 &  67.2 &  56.3 &  78.0 &  45.0\\
  7034.901  & 5.871 & --0.779 &  51.2 &       &  69.4 &       &       &      \\
  7680.266  & 5.863 & --0.609 &  68.5 & 100.7 &  85.6 &  80.0 &       &  75.2\\
  7760.628  & 6.206 & --1.356 &       &       &       &       &       &      \\
  8742.446  & 5.871 & --0.448 &  79.3 &       &  91.7 &       &       &      \\
  8892.720  & 5.984 & --0.705 &  62.0 &       &       &       &       &      \\
{\bf \ion{Ca}{i}}; $\log$ $\epsilon_{\odot}$(Ca)=6.36&&&&&&&&\\
  6166.439  & 2.521 & --1.074 & 103.3 & 112.2 & 145.7 & 131.4 & 163.0 & 124.9\\
  6455.598  & 2.523 & --1.350 &  93.1 & 106.7 & 135.6 & 123.2 & 154.5 & 116.0\\
  6499.650  & 2.523 & --0.839 & 128.2 & 134.4 & 171.9 & 161.6 &       & 158.1\\
{\bf \ion{Sc}{ii}}; $\log$ $\epsilon_{\odot}$(Sc)=3.10&&&&&&&&\\
  6320.851  & 1.500 & --1.747 &  21.9 &       &       &  33.1 &       &  24.9\\
{\bf \ion{Ti}{i}}; $\log$ $\epsilon_{\odot}$(Ti)=4.99&&&&&&&&\\
  5766.330  & 3.294 &   0.370 &  32.9 &  41.5 &  57.4 &  46.7 &  80.3 &  46.1\\
{\bf \ion{Cr}{i}}; $\log$ $\epsilon_{\odot}$(Cr)=5.67&&&&&&&&\\
  5787.965  & 3.323 & --0.138 &  66.6 &  85.6 & 109.8 & 100.7 & 136.8 &  90.7\\
  6882.475  & 3.438 & --0.238 &  56.7 &  75.1 &       &  90.9 &       &      \\
  6882.996  & 3.438 & --0.305 &  56.3 &  75.8 &       &       &       &  80.8\\
  6925.202  & 3.450 & --0.227 &  60.5 &  81.4 & 108.5 &  94.7 &       &  95.9\\
{\bf \ion{Fe}{i}}; $\log$ $\epsilon_{\odot}$(Fe)=7.67&&&&&&&&\\
  5543.937  & 4.218 & --1.155 &  74.4 & 101.1 & 113.5 & 104.9 & 136.2 &  92.8\\
  5638.262  & 4.221 & --0.882 &  90.3 & 118.9 & 134.9 & 121.7 &       & 108.7\\
  5679.025  & 4.652 & --0.863 &  68.3 &  90.7 & 100.4 &  92.8 & 117.9 &  84.0\\
  5732.275  & 4.992 & --1.473 &  25.3 &  45.1 &  40.2 &       &       &  26.1\\
  5806.717  & 4.608 & --0.984 &  63.2 &  91.2 &  97.9 &  89.8 & 118.8 &  76.7\\\hline
\end{tabular}
\end{center}
\end{table*}

\addtocounter{table}{-1}
\begin{table*}
\caption{Continued.}
\begin{center}
\begin{tabular}{lcrrrrrrr} \hline
  $\lambda$ (\AA) & $\chi$ (eV)  & log $gf$ & \multicolumn{6}{c}{EW (m\AA)$^a$}\\
     &    &          & UV For & VX Pyx & IN Vel & IS Vir & V764 Cen & V851 Cen\\\hline  
  5848.123  & 4.608 & --1.282 &  51.2 &  76.6 &  90.0 &  79.0 &  97.4 &  63.8\\
  5855.091  & 4.608 & --1.681 &       &  47.7 &  56.2 &  48.4 &  58.1 &  38.2\\
  5905.689  & 4.652 & --0.860 &  68.3 &  93.2 &  94.8 &  94.1 & 130.8 &      \\
  5909.970  & 3.211 & --2.731 &  65.5 &       & 102.5 &       & 117.4 &  71.4\\
  5927.786  & 4.652 & --1.243 &       &       &  76.1 &       &       &  60.1\\
  5929.667  & 4.549 & --1.332 &  47.2 &  74.7 &  75.5 &  74.2 &  91.4 &  60.3\\
  5930.173  & 4.652 & --0.347 &  98.5 & 126.4 & 131.8 & 126.3 &       &      \\
  5947.503  & 4.607 & --2.059 &       &  32.1 &  33.3 &       &       &      \\
  6078.491  & 4.796 & --0.414 &  86.0 & 115.7 & 124.7 & 114.2 & 148.1 &  99.1\\
  6078.999  & 4.652 & --1.123 &  58.1 &  82.4 &  94.2 &  86.4 & 105.0 &  76.7\\
  6094.364  & 4.652 & --1.749 &  25.7 &  49.4 &       &  46.9 &       &  35.0\\
  6098.280  & 4.559 & --1.940 &  24.6 &       &       &  45.1 &       &  34.3\\
  6151.617  & 2.176 & --3.486 &  83.1 &       & 126.1 & 123.2 &       & 106.3\\
  6165.361  & 4.143 & --1.645 &  54.9 &  80.5 &  91.7 &  82.0 & 112.0 &  67.7\\
  6187.987  & 3.944 & --1.740 &  66.1 &  91.6 & 102.2 &  94.5 & 116.0 &  79.5\\
  6219.279  & 2.198 & --2.551 & 134.7 & 161.3 &       &       &       & 160.4\\
  6252.554  & 2.404 & --1.867 & 161.5 &       &       &       &       &      \\
  6322.690  & 2.588 & --2.503 & 110.4 & 134.8 &       & 148.5 &       &      \\
  6335.328  & 2.198 & --2.432 & 143.2 & 167.8 &       &       &       & 168.1\\
  6336.823  & 3.687 & --0.896 & 131.3 & 161.7 &       & 172.1 &       & 161.6\\
  6436.411  & 4.187 & --2.538 &       &  38.1 &       &       &       &  23.7\\
  6469.213  & 4.835 & --0.774 &       &       &       &       &       &  79.2\\
  6593.871  & 2.433 & --2.342 & 126.0 &       & 179.4 & 168.2 &       & 145.5\\
  6699.162  & 4.593 & --2.172 &       &  32.8 &       &  25.7 &       &  19.0\\
  6713.771  & 4.796 & --1.606 &       &  48.7 &  50.4 &  43.4 &       &  32.2\\
  6725.353  & 4.104 & --2.370 &  27.8 &  48.2 &  54.4 &  47.1 &  56.0 &  33.5\\
  6726.661  & 4.607 & --1.200 &  53.1 &  78.8 &  83.3 &  78.2 & 105.2 &  64.8\\
  6733.151  & 4.639 & --1.594 &  34.7 &  57.4 &  61.4 &  56.6 &       &  44.1\\
  6745.090  & 4.580 & --2.192 &       &       &       &  30.5 &       &      \\
  6750.150  & 2.424 & --2.727 & 115.4 & 142.6 & 163.2 & 161.1 &       & 132.6\\
  6806.847  & 2.728 & --3.265 &  61.9 &  91.5 & 112.8 & 100.5 & 138.3 &  80.7\\
  6810.257  & 4.607 & --1.129 &  56.0 &  88.4 &  94.9 &  88.9 &       &  72.0\\
  6820.369  & 4.639 & --1.289 &  46.0 &  77.5 &  85.1 &  74.0 & 104.7 &      \\
  6843.648  & 4.549 & --0.934 &  66.8 &  98.3 & 102.6 &  95.7 &       &  76.9\\
  6857.243  & 4.076 & --2.203 &  30.7 &  54.5 &  63.1 &  57.6 &       &  41.6\\
  6862.492  & 4.559 & --1.509 &  37.9 &  65.1 &  69.4 &  68.3 &       &  49.9\\
  7022.953  & 4.191 & --1.184 &       &       & 116.2 &       &       &  98.8\\
  7219.678  & 4.076 & --1.715 &  59.0 &       &  90.0 &  88.8 &       &  71.2\\
  7306.556  & 4.178 & --1.684 &  53.5 &       &  95.3 &  79.3 &       &      \\
  7746.587  & 5.064 & --1.379 &  19.6 &       &       &  41.7 &       &  35.2\\
  7748.274  & 2.949 & --1.748 & 135.2 & 178.2 & 198.7 & 185.1 &       & 166.4\\\hline
\end{tabular}
\end{center}
\end{table*}

\addtocounter{table}{-1}
\begin{table*}
\caption{Continued.}
\begin{center}
\begin{tabular}{lcrrrrrrr} \hline
  $\lambda$ (\AA) & $\chi$ (eV)  & log $gf$ & \multicolumn{6}{c}{EW (m\AA)$^a$}\\
     &    &          & UV For & VX Pyx & IN Vel & IS Vir & V764 Cen & V851 Cen\\\hline  
  7751.137  & 4.992 & --0.841 &  52.4 &       &  93.1 &  86.8 &       &      \\
  7780.552  & 4.474 & --0.175 & 127.5 & 168.9 & 173.4 & 164.1 & 208.2 & 144.4\\
  7802.473  & 5.086 & --1.493 &  23.0 &  45.4 &  47.8 &  40.5 &       &      \\
  7807.952  & 4.992 & --0.602 &  64.1 & 101.4 & 105.1 &  88.5 &       &  80.6\\
  8922.643  & 4.992 & --1.570 &  25.6 &       &       &       &       &      \\
{\bf \ion{Fe}{ii}}; $\log$ $\epsilon_{\odot}$(Fe)=7.67&&&&&&&&\\
  5991.376  & 3.153 & --3.702 &  29.1 &  62.3 &       &  39.0 &       &  26.5\\
  6149.258  & 3.889 & --2.858 &  26.3 &  65.9 &  48.2 &  51.8 &  54.4 &  35.4\\
  6369.462  & 2.891 & --4.192 &       &       &       &  32.4 &       &      \\
  6416.919  & 3.892 & --2.750 &  38.0 &  64.4 &       &  60.6 &  51.7 &  40.0\\
  6456.383  & 3.904 & --2.209 &  54.6 &  96.1 &  71.5 &  78.4 &       &  50.9\\
  7711.723  & 3.904 & --2.625 &  36.0 &  80.3 &  54.8 &  59.1 &  54.8 &      \\
{\bf \ion{Co}{i}}; $\log$ $\epsilon_{\odot}$(Co)=4.92&&&&&&&&\\
  6454.990  & 3.632 & --0.233 &  33.6 &  48.6 &  57.4 &  47.2 &  65.1 &  32.7\\
{\bf \ion{Ni}{i}}; $\log$ $\epsilon_{\odot}$(Ni)=6.25&&&&&&&&\\
  5593.733  & 3.899 & --0.683 &  51.5 &  75.7 &  83.3 &  68.2 &  86.5 &  61.3\\
  5805.213  & 4.168 & --0.530 &  45.3 &  69.5 &  70.2 &  63.2 &  62.8 &  52.8\\
  6111.066  & 4.088 & --0.785 &  38.4 &  71.3 &  70.0 &  61.0 &       &  52.0\\
  6176.807  & 4.088 & --0.148 &  72.4 &  98.5 & 101.1 &  88.5 & 124.9 &  77.2\\
  6186.709  & 4.106 & --0.777 &  43.6 &  68.4 &  75.1 &       &       &      \\
  6204.600  & 4.088 & --1.060 &  30.8 &  54.1 &  56.4 &  46.6 &  70.3 &  32.5\\
  6223.981  & 4.106 & --0.876 &  33.6 &  61.2 &       &  55.8 &       &  46.1\\
  6772.313  & 3.658 & --0.890 &  62.5 &  87.2 &  95.0 &  86.2 & 101.2 &  72.4\\
  7555.598  & 3.848 &   0.069 & 105.9 & 142.1 & 144.6 & 131.2 &       & 112.4\\
  7797.586  & 3.899 & --0.144 &  89.1 & 121.8 & 122.4 & 118.7 & 147.2 & 105.2\\
{\bf \ion{Ba}{ii}}; $\log$ $\epsilon_{\odot}$(Ba)=2.13&&&&&&&&\\
  5853.668  & 0.604 & --0.758 &  92.8 & 132.4 & 142.2 & 151.5 & 165.0 & 119.9\\\hline
\end{tabular}
\begin{flushleft}
$^a$ A blank indicates that the EW was not reliably measurable for one of the following reasons: the line was affected by telluric features or cosmetic defaults (e.g., cosmic rays), the line was significantly van der Walls broadened (see Sect.~\ref{sect_methods}), or the Gaussian fit was judged unsatisfactory.
\end{flushleft}
\end{center}
\end{table*}

\end{document}